\documentclass[%
 aip,
 amsmath,amssymb,
 reprint,%
]{revtex4-2}

\usepackage{physics}
\usepackage{algpseudocode}

\usepackage{tikz}
\usepackage{subcaption}
\usepackage{graphicx}

\usepackage{epstopdf}

\usepackage{graphicx}
\usepackage{dcolumn}
\usepackage{bm}
\usepackage{comment} 
\usepackage[utf8]{inputenc}
\usepackage[T1]{fontenc}
\usepackage{mathptmx}
\usepackage{etoolbox}

\usepackage{xcolor}
\usepackage{tikz}
\usetikzlibrary{positioning, shapes.geometric, arrows.meta, calc, shadows, backgrounds, fit}

\definecolor{codegreen}{rgb}{0,0.6,0}
\definecolor{codegray}{rgb}{0.5,0.5,0.5}
\definecolor{codepurple}{rgb}{0.58,0,0.82}
\definecolor{backcolour}{rgb}{0.95,0.95,0.92}

\usepackage{amsmath}
\usepackage{amssymb}
\usepackage{graphicx}
\usepackage[margin=1in]{geometry}
\usepackage{hyperref}
\usepackage{listings}
\usepackage{xcolor}

\usepackage{graphicx}

\definecolor{codegreen}{rgb}{0,0.6,0}
\definecolor{codegray}{rgb}{0.5,0.5,0.5}
\definecolor{codepurple}{rgb}{0.58,0,0.82}
\definecolor{backcolour}{rgb}{0.95,0.95,0.92}

\lstdefinestyle{fortran_style}{
    backgroundcolor=\color{backcolour},   
    commentstyle=\color{codegreen},
    keywordstyle=\color{blue},
    numberstyle=\tiny\color{codegray},
    stringstyle=\color{codepurple},
    basicstyle=\ttfamily\footnotesize,
    breakatwhitespace=false,         
    breaklines=true,                 
    captionpos=b,                    
    keepspaces=true,                 
    language=Fortran,
    numbers=left,                    
    numbersep=5pt,                  
    showspaces=false,                
    showstringspaces=false,
    showtabs=false,                  
    tabsize=2
}
\makeatletter
\def\@email#1#2{%
 \endgroup
 \patchcmd{\titleblock@produce}
  {\frontmatter@RRAPformat}
  {\frontmatter@RRAPformat{\produce@RRAP{*#1\href{mailto:#2}{#2}}}\frontmatter@RRAPformat}
  {}{}
}%

\makeatother

\begin{document}

\preprint{AIP/123-QED}

\title[HyFrac.fun]{HyFrac.fun: A 3D Hydraulic Fracturing Simulator on Cloud}
\author{Jing Hu}
\thanks{Corresponding author: jhu@tongji.edu.cn.}
\affiliation{ 
	College of Civil Engineering, Tongji University
}%

\author{Qian Liu}%
\affiliation{ 
	School of Software, Shandong University
}%

\author{Jaroon Rungamornrat}
\affiliation{%
	Center of Excellence in Applied Mechanics and Structures, Department of Civil Engineering, Faculty of Engineering, Chulalongkorn University, Bangkok 10330, Thailand
}%
\affiliation{%
	GreenTech Nexus: Research Center for Sustainable Construction Innovation, Faculty of Engineering, Chulalongkorn University, Bangkok 10330, Thailand
}%



\date{\today}

\begin{abstract}
When multiple hydraulic fractures propagate simultaneously from a horizontal wellbore, elastic stress-shadow interactions generate complex non-planar three-dimensional geometries whose effect on subsequent reservoir drainage has infrequently been quantified, because the propagation and production solvers have historically been incompatible stand-alone tools. This paper presents HyFrac.fun, a cloud-native platform that bridges this gap by exploiting a structural isomorphism between the two SGBEM--FEM governing operator systems. The platform enables automated zero-conversion handoff of the evolved 3D fracture mesh directly to the steady-state Darcy production solver for realizing a fully integrated lifecycle simulation of multi-stage non-planar hydraulic fractures. The lifecycle analysis reveals a double shadow phenomenon: the mechanical stress shadow that suppresses inner-fracture growth during stimulation mirrors a fluid pressure shadow that reduces the inner fracture's drawout rate at small cluster spacing. Critically, switching to a shear-thinning power-law fracturing fluid leaves the fracture trajectories and production rates almost unchanged, demonstrating that stress-shadow-controlled fracture geometry instead of fluid rheology is the primary determinant of long-term production efficiency at equal injection rates. These physics findings are accessible from integrated fracture propagation and production simulations.
\end{abstract}

\maketitle

\section{Introduction}

Hydraulic fracturing has become a foundational technique for unlocking unconventional oil and gas resources \cite{Montgomery2010, Gallegos2015, Clark1949, Hubbert1957, Valko1995}. By injecting high-pressure fluids, fractures are created and propagated, increasing the effective permeability of tight formations and allowing hydrocarbons to reach production wells \cite{Economides2000, Yew1997, Warpinski1987, Daneshy1973}. The geometry and connectivity of these induced fractures govern the efficiency of stimulation treatments and the long-term productivity of the reservoir \cite{Cipolla2008, Mayerhofer2010, Olson2008}. Accurate modelling of both the propagation of fractures and the subsequent production performance is therefore critical for optimal well design and reservoir management \cite{Adachi2007, Weng2011, omidi2015adaptive}.

Early numerical models of hydraulic fracturing were primarily established through two-dimensional plane strain formulations, most notably the Khristianovic-Geertsma-de Klerk (KGD) \cite{Geertsma1969, Khristianovic1955} and Perkins-Kern-Nordgren (PKN) models \cite{Perkins1961, Nordgren1972}. The KGD model assumes a state of plane strain in the horizontal plane, making it suitable for short fractures where the length-to-height ratio is small, whereas the PKN model assumes plane strain in the vertical plane, providing a more appropriate description for elongated fractures confined within a fixed-height pay zone \cite{Valko1995, Adachi2007}. As the industry demanded more flexible tools capable of handling multilayered reservoirs, pseudo-three-dimensional (P3D) approaches were developed in the 1980s as a computational compromise \cite{settari1986development, Cleary1980}. These models, introduced by pioneers such as Settari and Cleary (1986)\cite{settari1986development} and Palmer and Carroll (1983)\cite{palmer1983numerical}, extended the 1D diffusion-type equations of the PKN framework by allowing the fracture height to vary dynamically in response to stress barriers and rock toughness \cite{Meyer1986}. While P3D models are computationally efficient and suitable for scenarios where stress barriers confine fractures to relatively simple shapes, they are inherently limited by their reliance on local elastic compliance and the assumption of a predefined and often elliptical geometry \cite{Rungamornrat2005}.

To simulate non-planar fracture propagation, high-fidelity numerical techniques have been developed that couple rock elasticity and fluid flow in three dimensions \cite{Lecampion2018, Gordeliy2013, miehe2010phase, Bourdin2008}. A prominent approach is the coupling of symmetric Galerkin boundary element methods (SGBEM) \cite{Bonnet1998, Phan2003,sutradhar2008symmetric} with finite-element formulations for channel flow \cite{Rungamornrat2005,  Ganis2014}. In this approach, the elasticity problem is formulated on the fracture surface and solved via a weakly singular boundary integral equation, while fluid flow inside the fracture is treated using finite elements \cite{Rungamornrat2005}. Recently, Hu and Mear extended this approach to steady-state production analysis by coupling Darcy flow in the reservoir to channel flow within the fracture \cite{Hu2022, Hu2025}. These SGBEM–FEM methods can model complex fracture geometries and accommodate shear-thinning or shear-thickening fluids, but they remain computationally intensive \cite{Rungamornrat2005, Hu2022, Lecampion2018}.

Although its computational cost is significantly reduced compared to domain based approaches, the computational burden of these SGBEM-FEM models is significant stemming from two primary sources \cite{Rungamornrat2005, hu2024efficient}. First, the SGBEM formulation for elasticity results in a dense, symmetric stiffness matrix, where the computational cost of formation scales quadratically $\mathcal{O}(N_{el}^2)$, with the number of elements $N$ \cite{Bonnet1998, Phan2003, hu2024efficient}. For a dynamic propagation simulation, this computationally expensive matrix must be re-formed and re-solved at every time step \cite{Rungamornrat2005}. Second, the fracture surface itself is evolving, which necessitates a robust and complex adaptive remeshing algorithm \cite{ Rungamornrat2005, omidi2015adaptive}. This geometric engine must dynamically manage the mesh topology as the crack advances, by performing element splitting in regions of high growth and element coarsening (merging) in the fracture interior all while maintaining high element quality to ensure numerical accuracy \cite{omidi2015adaptive}.

These dual challenges, i.e. complex remeshing logic and the need for highly optimized matrix assembly, are the primary bottlenecks in 3D fracture simulation \cite{hu2024efficient, Rungamornrat2005}. To be computationally tractable, the simulator's core routines must be aggressively optimized \cite{hu2024efficient}. This includes algorithmic optimizations, such as incremental matrix updates that exploit temporal locality by caching the interactions of stationary elements, as well as high-performance computing techniques \cite{hu2024efficient}. Specifically, scalable shared-memory parallelism via OpenMP is not merely an option but an essential requirement to distribute the $\mathcal{O}(N_{el}^2)$ matrix formation workload across all available CPU cores, making the problem feasible in a reasonable timeframe \cite{Dagum1998, chandra2001parallel}. These intense computational and algorithmic requirements compound the practical barriers to adoption. Many existing simulators are written in legacy languages, require specialized compilation, and depend on local high-performance computing (HPC) hardware \cite{Zhao2007}. Furthermore, post-processing and visualizing the large, complex 3D datasets typically involve proprietary software, such as Tecplot 360 \cite{heidbach2020manual}, which may not be available to all users. Consequently, these sophisticated modelling tools remain largely confined to large service companies and academic researchers, while engineers in smaller organisations frequently rely on simplified or commercial packages \cite{Cipolla2008}. 

Beyond accessibility, a critical and largely unresolved gap exists in the simulation workflow itself. The fracture propagation solvers \cite{Rungamornrat2005, hu2024efficient} and the production analysis solvers \cite{Hu2022, Hu2025} have historically been developed and operated as independent tools. Transitioning from a completed propagation simulation, whose final geometry may consist of a large amount of adaptively remeshed non-planar surface elements, to a production analysis requires manual mesh extraction, format conversion, and re-parameterization, which is a process that is error-prone, time-consuming, and a practical obstacle to systematic design iteration. Few existing open tool provides an automated end-to-end pipeline connecting three-dimensional non-planar fracture propagation directly to steady-state production analysis on the same geometry. This paper addresses this gap by demonstrating that the propagation and production formulations share a structurally identical operator framework, a property that makes automated lifecycle integration both mathematically natural and implementationally straightforward. This workflow gap is not merely a software inconvenience, while it represents a fundamental barrier to understanding the physics of coupled stimulation and production. When fractures propagate under stress-shadow interactions, their non-planar three-dimensional geometries emerge as the result of a complex fluid--solid coupling. These geometries directly determine the subsequent reservoir drainage pattern, yet their effect on production has rarely been quantified because few tool could carry the evolved 3D mesh directly into a production solver.

Cloud computing offers a potential solution to both the accessibility and the computational scaling problems \cite{Armbrust2010}. Cloud platforms provide on-demand access to powerful, multi-core computational resources necessary for parallelized solvers, along with scalable storage and high-speed networking, allowing developers to offload heavy computations from user machines \cite{Buyya2009, Foster2008}. Combined with modern web technologies, cloud services can deliver interactive simulation workflows via standard browsers, enabling users to submit jobs, monitor progress and visualise results remotely \cite{ Zhao2007}. By adopting a software-as-a-service (SaaS) model, it becomes possible to democratise access to advanced numerical simulations and reduce the overhead associated with software installation, licensing and hardware maintenance \cite{ Benlian2010}.

This paper presents HyFrac.fun, a fully cloud-deployed platform that integrates these high-performance, parallelized, and adaptively remeshed SGBEM–FEM engines for simulating the entire lifecycle of a hydraulic fracture. Advancing from validated solvers \cite{Rungamornrat2005, Hu2022}, HyFrac.fun allows users to configure simulations through a web interface, specifying fracture counts, spacing, rock properties, fluid rheology and leak-off models. A FastAPI backend orchestrates the workflow \cite{lubanovic2023fastapi}: it generates input files, schedules computations on cloud hardware, monitors execution, and stores outputs. After simulation, ParaView/VTK renders the results server-side \cite{Ahrens2005, Schroeder2006}, and the TRAME framework serves interactive 3D visualisations to the user’s browser \cite{Trame2021}.

The integration of these components is intended to bridge the gap between cutting-edge research and practical engineering applications. By decoupling the computational workload from the user’s local machine and hiding the complexity of code compilation and execution behind a web interface, HyFrac.fun enables engineers with modest computing resources to perform sophisticated simulations \cite{hu2024efficient}. Containerisation and orchestration technologies ensure reproducibility and simplify maintenance, as software dependencies are managed centrally. The end-to-end workflow, from input submission to result exploration, promotes collaboration and transparency, addressing common challenges in simulation-driven decision making.

The remainder of this paper is structured as follows. Section \ref{sec:formulations} details the governing SGBEM-FEM numerical formulations for both fracture propagation and well production, as well as the dynamic adaptive remeshing procedure. Section \ref{sec:para} focuses on computational performance, outlining the incremental matrix updates and OpenMP parallelism strategies essential to the simulator's efficiency. Section \ref{sec:cloud_arch} presents the cloud-native software architecture, encompassing the web front-end, backend orchestration, and remote visualization service. Following this, Section \ref{sec:examples} provides numerical verifications and complex multi-stage fracture simulation examples. Finally, Section \ref{sec:conclusions} offers concluding remarks.

\section{Numerical Formulations}\label{sec:formulations}

HyFrac.fun integrates two simulation engines whose formulations are adopted from prior works by the authors: the hydraulic fracture propagation engine follows Rungamornrat et al. (2005)~\cite{Rungamornrat2005} and the steady-state well production engine follows Hu and Mear (2022)~\cite{Hu2022}. Both formulations are reviewed here for completeness and to expose a structural property that is central to this platform's design.

Despite modeling fundamentally different physical regimes, which are a transient nonlinear problem on an evolving fracture geometry versus a steady-state linear problem on a fixed geometry, the governing weak-form equations for both phases can be expressed in a structurally isomorphic abstract operator framework. As detailed subsequently in Section~\ref{sec:unified_operators}, this framework utilizes a unified set of bilinear and linear operators to represent rock stiffness, fluid conductance, and solid-fluid coupling. This isomorphism is the mathematical property that enables HyFrac.fun's automated, zero-conversion handoff of the evolving 3D fracture mesh from the propagation engine to the production engine.

\subsection{Hydraulic Fracture Propagation Model}

The simulation of hydraulic fracture propagation is governed by the coupled mechanics of rock deformation and fluid flow. The non-planar fracture evolution  in three-dimensional, homogeneous, and linearly elastic media is solved with a symmetric Galerkin boundary element method, while the channel flow in the fracture is solved using a standard Galerkin finite element method.

\subsubsection{Linear Elastic Fracture Mechanics}
\begin{figure}[h!]
	\centering
	\includegraphics[trim=1.1cm 1.3cm 1.1cm 1.1cm, clip, width=0.43\textwidth]{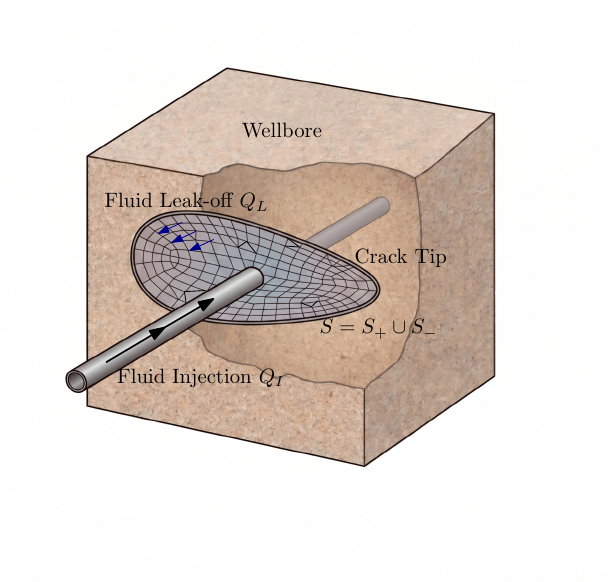}
	\caption{Schematic of a fracture embedded in a rock matrix.}
	\label{fig:matrix_permute}
\end{figure}
The elastic response of the rock mass to fracture opening is modeled using a weak-form traction boundary integral equation, which avoids the need to mesh the entire 3D domain. For a fracture surface $S= S_+ \cup S_-$ (see Figure \ref{fig:matrix_permute}), the relationship between the traction $\mathbf{t}$ on the surface and the relative crack-face displacement (opening) $\Delta\mathbf{u}$ is 
\begin{equation}
\begin{aligned}\label{eq:fracbie}
\int_{S_+}D_{l}\Delta\tilde{u}_{k}(x)\int_{S_+}C_{mj}^{lk}(y-x)D_{m}\Delta u_{j}(y)dS(y)dS(x)=\\-\int_{S_+} t_{k}(x)\Delta\tilde{u}_{k}(x)dS(x) 
\end{aligned}
\end{equation}
where $\Delta\tilde{\mathbf{u}}$ is a test function for the displacement jump, $D_m$ is a surface differential operator, and $C_{mj}^{lk}$ is a weakly-singular kernel of order $\mathcal{O}(1/r)$, where $r = \|y-x\|$. For an isotropic elastic medium with shear modulus $\mu$ and Poisson's ratio $\nu$, the kernel simplifies to
\begin{equation}
\begin{aligned}
C_{mj}^{tk}(y-x)=\frac{\mu}{4\pi(1-\nu)r}\Big( (1-\nu) \delta_{tk}\delta_{mj}+2\nu\delta_{mk}\delta_{jt} \\- \delta_{jk}\delta_{mt} - \frac{(y_{j}-x_{j})(y_{k}-x_{k})}{r^{2}}\delta_{mr}\Big)
\end{aligned}
\end{equation}

This SGBEM formulation is advantageous as the weak singularity allows for the use of standard $C^0$ continuous elements for discretization, avoiding the complexities associated with the hypersingular integrals in collocation BEM. Along the fracture front, special crack-tip elements are employed to accurately capture the characteristic square-root singularity of the displacement field, enabling the direct and accurate computation of stress intensity factors ($K_I, K_{II}, K_{III}$).

\subsubsection{Fluid Flow Formulation}

The flow of injected fluid within the evolving non-planar fracture is modeled as channel flow. The governing equations, which account for mass conservation and a Newtonian or power-law non-Newtonian fluid rheology, are cast into a weak form suitable for a Galerkin FEM treatment:

\begin{equation}
	\int_{S_+}\frac{w^{3}}{12\eta}(D_{m}\tilde{p})(D_{m}p)dS=\int_{S_+}\tilde{p}\left[Q_{I}-Q_{L}-\frac{\partial w}{\partial t}\right]dS
\end{equation}
where $p$ is the fluid pressure, $\tilde{p}$ is a pressure test function, $w$ is the fracture width, $Q_I$ is the fluid injection rate per unit area, and $Q_L$ is the fluid leak-off rate into the formation.

It is important to clarify that during the transient fracture propagation phase, the interaction between the reservoir flow and the fracture flow is not modeled directly via a fully coupled 3D transient Darcy domain. Instead, this fluid exchange is efficiently approximated through the leak-off term $Q_L$. This treatment is adopted because the physical timescale of fracture propagation is extremely short, and simulating the fully coupled 3D transient porous media flow simultaneously with the rapidly evolving fracture geometry is computationally prohibitive. Thus, the local fluid invasion is approximated as a transient diffusion process into the matrix using Carter's leak-off model \cite{carter1957derivation}:
\begin{equation}
	Q_{L}=\frac{2c_{L}}{\sqrt{t-t_{S}(x)}}
\end{equation}
where $c_L$ is the leak-off coefficient and $t_S(x)$ is the time at which the fracture front exposes the formation at point $x$. This contrasts with the well-production model discussed later (Section \ref{sec:wellcouple}), where the system operates in a steady-state regime and long-term reservoir-wide pressure gradients dictate fluid transport, making direct coupling between the Darcy reservoir flow and the fracture channel flow both physically necessary and computationally tractable.

The effective viscosity $\eta$ for a power-law fluid is a non-linear function of the fracture width and the local pressure gradient:
\begin{equation}
	\eta=\frac{2n+1}{6n}(2K)^{\frac{1}{n}}\|w\nabla p\|^{1-\frac{1}{n}}
\end{equation}
where $n$ is the power-law exponent and $K$ is the consistency coefficient. Note that the above power-law constitutive model reduces to a Newtonian fluid when $n=1$, whose viscosity is $\eta=K$.

\subsubsection{Coupling Equations for Hydraulic Fracture Propagation}
\label{sec:operators}

The numerical solution to the hydraulic fracturing problem requires the monolithic coupling of the solid mechanics and fluid flow models. The two systems are physically linked through two primary conditions. First, the total traction $\mathbf{t}$ acting on the fracture faces is a superposition of the fluid pressure $p$ and the in-situ stress $\boldsymbol{\sigma}^o$:
\begin{equation}
\mathbf{t} = -p\mathbf{n} - \boldsymbol{\sigma}^o\mathbf{n}
\end{equation}
Second, the fracture opening width $w$, which dictates the hydraulic conductivity of the fluid channel, is defined by the normal component of the relative crack-face displacement jump $\Delta\mathbf{u}$:
\begin{equation}
w = \Delta\mathbf{u} \cdot \mathbf{n}
\end{equation}
To facilitate a robust numerical implementation, the governing weak-form equations are expressed in a compact operator form, separating the system into a series of bilinear and linear operators that correspond to the stiffness, coupling, and forcing terms of the final discrete system. The continuous weak-form system is written as:
\begin{align} 
    \mathcal{A}(\Delta\tilde{\mathbf{u}}, \Delta\mathbf{u}) + \mathcal{B}(\Delta\tilde{\mathbf{u}}, p) &= \mathcal{R}_c(\Delta\tilde{\mathbf{u}}; \boldsymbol{\sigma}^o) \label{eq:compact_solid} \\
    \mathcal{B}\left(\frac{\partial \Delta\mathbf{u}}{\partial t}, \tilde{p}\right) + \mathcal{C}(\tilde{p}, p) &= \mathcal{R}_f(\tilde{p}; Q_I, Q_L) \label{eq:compact_fluid}
\end{align}
The elasticity bilinear operator $\mathcal{A}$ represents the elastic stiffness of the rock matrix, arising from the left-hand side of the traction integral equation. It is a symmetric positive-definite operator defined by the double surface integral:
\begin{equation}
\begin{aligned}
    \mathcal{A}(\Delta\tilde{\mathbf{u}}, \Delta\mathbf{u}) = \int_{S_+}\int_{S_+} \left[D_l \Delta\tilde{u}_k(y)\right]\\ C_{mj}^{lk}(\xi-y) \left[D_m \Delta u_j(\xi)\right] dS(\xi)dS(y)
\end{aligned}
\end{equation}
The coupling bilinear operator $\mathcal{B}$ represents the interaction between the solid mechanics and fluid flow, appearing in two forms corresponding to the work done by fluid pressure and the rate of change of fracture volume:
\begin{align}
    \mathcal{B}(\Delta\tilde{\mathbf{u}}, p) &= -\int_{S_+} \Delta\tilde{u}_k(y) p(y) n_k(y) dS(y) \\
    \mathcal{B}\left(\frac{\partial \Delta\mathbf{u}}{\partial t}, \tilde{p}\right) &= -\int_{S_+} \tilde{p}(y) \frac{\partial w}{\partial t} dS(y)
\end{align}
The fluid flow bilinear operator $\mathcal{C}$ represents the hydraulic conductance of the fracture. It is non-linear due to its dependence on the fracture width $w^3$ and the effective viscosity $\eta$:
\begin{equation}
    \mathcal{C}(\tilde{p}, p) = \int_{S_+} \frac{w^3}{12\eta} (D_m \tilde{p}) (D_m p) dS(x)
\end{equation}
The forcing terms are the solid mechanics linear operator $\mathcal{R}_c$, representing the work done by the in-situ stress, and the fluid flow linear operator $\mathcal{R}_f$, representing the net volumetric flux:
\begin{align}
    \mathcal{R}_c(\Delta\tilde{\mathbf{u}}; \boldsymbol{\sigma}^o) &= \int_{S_+} \Delta\tilde{u}_k(y) \sigma_{lk}^o(y) n_l(y) dS(y) \\
    \mathcal{R}_f(\tilde{p}; Q_I, Q_L) &= \int_{S_+} \tilde{p}(y) \left[ Q_I(y) - Q_L(y) \right] dS(y)
\end{align}
After spatial discretization with isoparametric elements and temporal discretization with a backward finite difference scheme, the continuous physical operators are instantiated into specialized matrix data structures. The fully coupled, non-linear system of algebraic equations at time step $t^{i+1}$ takes the matrix form:
\begin{equation}
\begin{bmatrix} \mathbf{A} & \mathbf{B} \\ \mathbf{B}^{T} & \mathbf{C}(\mathbf{w}^{i+1}, \nabla p^{i+1}) \end{bmatrix} \begin{bmatrix} \Delta\mathbf{U}^{i+1} \\ \mathbf{P}^{i+1} \end{bmatrix} = \begin{bmatrix} \mathbf{R}_c \\ \mathbf{R}_f^* \end{bmatrix}
\label{eq:matrix_system}
\end{equation}
where $\Delta\mathbf{U}$ and $\mathbf{P}$ are the nodal vectors of relative displacement and fluid pressure, respectively, and $\mathbf{R}_f^*$ includes historical terms from the previous time step. The mathematical differences between the abstract component operators strictly govern their computer-memory implementations: the double-area integration of the boundary element operator $\mathcal{A}$ inherently generates a dense and symmetric $\mathcal{O}(N_{el}^2)$ stiffness matrix $\mathbf{A}$, whereas the localized spatial derivatives of operators $\mathcal{B}$ and $\mathcal{C}$ produce sparsely populated $\mathcal{O}(N_{el})$ finite element matrices. Furthermore, the overall system's non-linearity is concentrated entirely in the fluid conductance matrix $\mathbf{C}$, which relies heavily on the evolving unknown fields $(\Delta\mathbf{U}^{i+1}, \mathbf{P}^{i+1})$.

This system is solved monolithically at each time step using a Newton-Raphson iterative scheme. Let the solution at iteration $k$ be $(\Delta\mathbf{U}^{(k)}, \mathbf{P}^{(k)})$. The system is linearized to solve for the increments $(\delta\Delta\mathbf{U}^{(k+1)}, \delta\mathbf{P}^{(k+1)})$ as
\begin{equation}
\begin{bmatrix} \mathbf{A} & \mathbf{B} \\ \mathbf{B}^{T} + \frac{\partial(\mathbf{C}\mathbf{P})}{\partial \Delta\mathbf{U}} & \mathbf{C} + \frac{\partial(\mathbf{C}\mathbf{P})}{\partial \mathbf{P}} \end{bmatrix}^{(k)} \begin{bmatrix} \delta\Delta\mathbf{U}^{(k+1)} \\ \delta\mathbf{P}^{(k+1)} \end{bmatrix} = \begin{bmatrix} \mathbf{r}_c^{(k)} \\ \mathbf{r}_f^{(k)} \end{bmatrix}
\end{equation}
where the right-hand side vectors are the residuals at iteration $k$:
\begin{align}
    \mathbf{r}_c^{(k)} &= \mathbf{R}_c - \mathbf{A}\Delta\mathbf{U}^{(k)} - \mathbf{B}\mathbf{P}^{(k)} \\
    \mathbf{r}_f^{(k)} &= \mathbf{R}_f^* - \mathbf{B}^T\Delta\mathbf{U}^{(k)} - \mathbf{C}^{(k)}\mathbf{P}^{(k)}
\end{align}
The Jacobian matrix includes the derivatives of the non-linear fluid term with respect to the primary variables. The solution is then updated: $\Delta\mathbf{U}^{(k+1)} = \Delta\mathbf{U}^{(k)} + \delta\Delta\mathbf{U}^{(k+1)}$ and $\mathbf{P}^{(k+1)} = \mathbf{P}^{(k)} + \delta\mathbf{P}^{(k+1)}$. This iterative process continues until the L2-norm of the incremental update falls below a specified tolerance.


In the implementation, the assembly of these matrices is a critical and computationally intensive step. Dedicated routines orchestrate the calculation of the dense solid stiffness matrix $\mathbf{A}$ by looping over all pairs of elements, while the coupling matrix $\mathbf{B}$ is assembled in a separate procedure. To handle the significant computational cost, these high-level routines are parallelized using OpenMP directives \cite{Dagum1998}. During the parallel summation of local element contributions, race conditions are prevented by enclosing the global matrix update operations within appropriate synchronization constructs (see Section.~\ref{sec:para}). Once the system is converged, fracture growth is determined by a mixed-mode stress intensity factor criterion, and the mesh is updated via the adaptive remeshing algorithm before proceeding to the next time step.

\subsection{Well Production Model}

The following formulation for steady-state production is adopted from Hu and Mear (2022)~\cite{Hu2022}, which developed the SGBEM--FEM coupling for steady-state Darcy flow from a porous reservoir matrix into a fracture network. The present work advances this framework by applying it to evolved non-planar fracture geometries generated by actual multi-stage propagation simulations, enabling quantitative lifecycle analysis of fractures whose shapes are determined by physical stress-shadow interactions rather than geometric assumptions. After the propagation simulation terminates, the final evolved fracture mesh is passed directly to this production engine without manual conversion.

\subsubsection{Reservoir Flow (SGBEM)}

The steady-state flow through a homogeneous, isotropic porous reservoir is governed by Darcy's law, which leads to the Laplace equation for pressure. A weakly-singular pressure integral equation is established for the fracture surfaces, relating the pressure $p$ on the fracture to the sum of the fluid flux $\Sigma q$ entering the fracture from the matrix:
\begin{equation}
\begin{aligned}\label{eq:wellbie}
\int_{S^{+}}\Sigma\tilde{q}(x)p(x)dS(x) - \int_{S^{+}}\Sigma\tilde{q}(x)p_{0}dS(x) = \\-\int_{S^{+}}\Sigma\tilde{q}(x)\int_{S^{+}}P(\xi,x)\Sigma q(\xi)dS(\xi)dS(x)
\end{aligned}
\end{equation}
where $\Sigma\tilde{q}$ is a test function for the flux, $p_0$ is the far-field reservoir pressure, and $P(\xi, x)$ is the fundamental solution for the pressure field, given by
\begin{equation}
P(\xi,x)=\frac{\mu_{res}}{4\pi\kappa r}
\end{equation}
where $\mu_{res}$ is the reservoir fluid viscosity and $\kappa$ is the medium permeability. A special crack-tip element is used to capture the $\mathcal{O}(1/\sqrt{r})$ asymptotic behavior of the Darcy flux near the fracture perimeter.

\subsubsection{Fracture Flow (FEM)}

Flow within the static fracture geometry is again modeled as channel flow. This formulation contrasts with the fracture propagation model by treating the fracture domain as physically static and the fluid as a constant-viscosity Newtonian medium under steady-state conditions. For steady-state production of a Newtonian fluid, the weak-form equation simplifies to
\begin{equation}
\begin{aligned}
-\int_{S^{+}}\frac{w^{3}(x)}{12\mu}D_{m}\tilde{p}(x)D_{m}p(x)dS(x) = \\
\int_{S^{+}}\tilde{p}(x)[Q_{in}(x)-Q_{o}(x)]dS(x)
\end{aligned}
\end{equation}
where $\mu$ is the constant produced fluid viscosity, $Q_{in}$ is the fluid infiltration rate from the matrix to the fracture, and $Q_o$ is the fluid draw-out rate at the wellbore.

\subsubsection{Coupling Equations for Well Production}\label{sec:wellcouple}


The simulation of well production requires coupling the Darcy flow in the reservoir matrix with the channel flow inside the established fracture network. The two systems are physically linked at the fracture surface $S^{+}$ by the condition of mass conservation: the flux $\Sigma q$ entering the fracture from the reservoir must equal the flux $Q_{in}$ infiltrating the fracture channel:
\begin{equation}
Q_{in}(x) = \Sigma q(x)
\end{equation}
To facilitate a robust numerical implementation, the governing weak-form equations are expressed in a compact operator form, separating the system into a series of bilinear and linear operators. The continuous weak-form system is written as
\begin{align}
\mathcal{A}(\Sigma\tilde{q},\Sigma q)+\mathcal{B}(\Sigma\tilde{q},p) &= \mathcal{R}_{q}(\Sigma\tilde{q};p_{0}) \label{eq:compact_darcy}\\
\mathcal{B}^{T}(\tilde{p},\Sigma q)+\mathcal{C}(\tilde{p},p) &= \mathcal{R}_{p}(\tilde{p};Q_{o})\label{eq:compact_channel}
\end{align}
The reservoir flow bilinear operator $\mathcal{A}$ represents the pressure response in the matrix due to the flux distribution, arising from the BEM formulation for Darcy flow:
\begin{equation}
\mathcal{A}(\Sigma\tilde{q},\Sigma q) = \int_{S^{+}}\Sigma\tilde{q}(x)\int_{S^{+}}P(\xi,x)\Sigma q(x)dS(\xi)dS(x)
\end{equation}
The coupling bilinear operator $\mathcal{B}$ represents the coupling between the matrix flux and the fracture pressure:
\begin{equation}
\mathcal{B}(\Sigma\tilde{q},p) = \int_{S^{+}}\Sigma\tilde{q}(x)p(x)dS(x)
\end{equation}
The fracture flow bilinear operator $\mathcal{C}$ represents the hydraulic conductance of the fracture channel, derived from the FEM formulation:
\begin{equation}
\mathcal{C}(\tilde{p},p) = -\int_{S^{+}}\frac{w^{3}(x)}{12\mu}D_{m}\tilde{p}(x)D_{m}p(x)dS(x)
\end{equation}
The forcing terms are the reservoir load linear operator $\mathcal{R}_q$, representing the load from the far-field reservoir pressure $p_0$, and the wellbore draw-out linear operator $\mathcal{R}_p$, representing the fluid sink $Q_o$ at the wellbore:
\begin{align}
\mathcal{R}_{q}(\Sigma\tilde{q};p_{0}) &= \int_{S^{+}}\Sigma\tilde{q}(x) p_0 dS(x) \\
\mathcal{R}_{p}(\tilde{p};Q_{o}) &= \int_{S^{+}}\tilde{p}(x) Q_o(x) dS(x)
\end{align}
This system of operator equations is discretized using a standard Galerkin procedure. The continuous unknown fields $p(x)$ and $\Sigma q(x)$ are approximated using a set of shape functions $\mathbf{\psi}(x)$, such that $p(x) \approx \mathbf{\psi}(x) \mathbf{P}$ and $\Sigma q(x) \approx \mathbf{\psi}(x) \mathbf{Q}$. The operator equations are then tested against each basis function, yielding a system of linear algebraic equations. This system can be expressed in the following block matrix form:
\begin{equation}
\begin{bmatrix} 
\mathbf{A} & \mathbf{B} \\ 
\mathbf{B}^{T} & \mathbf{C} 
\end{bmatrix} 
\begin{bmatrix} 
\mathbf{Q} \\ 
\mathbf{P} 
\end{bmatrix} 
= 
\begin{bmatrix} 
\mathbf{R}_q \\ 
\mathbf{R}_p 
\end{bmatrix}
\end{equation}
where $\mathbf{Q}$ and $\mathbf{P}$ are the global vectors of nodal unknowns for flux and pressure, respectively. The matrices $\mathbf{A}$, $\mathbf{B}$, and $\mathbf{C}$ are the discrete matrix representations of their corresponding operators, and $\mathbf{R}_q$ and $\mathbf{R}_p$ are the global load vectors. Unlike the non-linear transient propagation problem, this formulation results in a symmetric linear system of equations which is solved directly to obtain the steady-state nodal pressure and flux distributions across the fracture network.

\subsection{Unified Operator Framework and Automated Lifecycle Handoff}
\label{sec:unified_operators}

A key structural observation motivating the integrated design of HyFrac.fun is that, despite their physical differences, both the hydraulic fracture propagation system (Eqs.~\ref{eq:compact_solid}--\ref{eq:compact_fluid}) and the steady-state production system (Eqs.~\ref{eq:compact_darcy}--\ref{eq:compact_channel}) can be expressed in the same abstract operator form. Setting the operators side by side can make the isomorphism explicit.

In both systems, the operator $\mathcal{A}$ is a dense symmetric double-surface-integral operator arising from the boundary element formulation: in the fracture propagation engine it encodes the elastic stiffness of the rock matrix via the weakly-singular traction integral equation, while in the production engine it encodes the Darcy pressure response of the porous reservoir. The operators $\mathcal{B}$ and $\mathcal{C}$ are sparse localized finite element operators in both cases: $\mathcal{B}$ couples the exterior/interior degrees of freedom and $\mathcal{C}$ governs the hydraulic conductance of the fracture channel. The physical interpretation of each operator changes between phases, but their mathematical structure which consists of a dense BEM operator $\mathcal{A}$of $\mathcal{O}(N_{el}^2)$ assembly complexity and sparse FEM operators $\mathcal{B}$ and $\mathcal{C}$ of $\mathcal{O}(N_{el})$ assembly complexity remains identical.

This structural isomorphism has two important consequences for the platform: 1. Shared computational infrastructure: the incremental stiffness update strategy, cache-optimized element-renumbering permutation, and hybrid OpenMP parallelization described in Section~\ref{sec:para} are applicable to both engines without modification, since both assemble the same type of packed dense matrix for $\mathcal{A}$ and sparse matrices for $\mathcal{B}$ and $\mathcal{C}$; 2. Automated mesh and field-state handoff: after the propagation simulation terminates, the complete state of the solver, i.e. the adaptively remeshed fracture surface, the nodal width field $w$, the pressure field $p$, and the element connectivity arrays, is transferred directly to the production engine. No intermediate geometry conversion or re-meshing is required, because both engines operate on the same mesh topology and nodal data structures. The HyFrac.fun backend (Section~\ref{sec:cloud_arch}) implements this handoff programmatically: the final VTU output of the propagation engine is loaded as the geometry of the production engine in a single automated pipeline step. This eliminates the manual post-processing bottleneck that, in existing workflows, prevents routine lifecycle analysis in practice. This represents an early endeavor for an automated end-to-end integration of three-dimensional non-planar multi-stage hydraulic fracture propagation with steady-state Darcy production analysis on the resulting evolved geometry.

\subsection{Adaptive Remeshing for Fracture Propagation}
\label{sec:remeshing}

The prior work that employ similar adaptive meshing algorithms~\cite{Rungamornrat2005,hu2024efficient} focus on the mathematical formulation and numerical results, and do not provide sufficient detail on the algorithmic implementation for independent reproduction. The present subsection addresses this gap by presenting the full implementation: the four-case element splitting and coarsening decision tree, the characteristic-length-driven quality control criterion, and the localized solution-field projection with singularity regularization for newly created tip nodes are each described in sufficient detail to serve as a self-contained implementation reference. The simulation of fracture propagation necessitates a dynamic meshing strategy capable of conforming to the evolving geometry of the crack surface. A static mesh would quickly become invalid as the fracture advances, leading to severe element distortion and a degradation of numerical accuracy. The numerical engine employs an adaptive remeshing algorithm at the conclusion of each converged time step to update the mesh topology and project the solution state onto the new configuration. This process is fundamental to the simulator's ability to model long-duration fracture growth and complex non-planar trajectories. The procedure can be conceptually divided into three main stages: fracture front advancement, adaptive element management, and solution data projection.

\subsubsection{Fracture Front Advancement}

The impetus for remeshing is the physical advancement of the fracture front, which is governed by Linear Elastic Fracture Mechanics (LEFM). Following the convergence of the coupled solid-fluid system for a given time step, the mixed-mode stress intensity factors, i.e. $K_I$ and $K_{II}$, are computed at each node along the fracture perimeter. These factors dictate the subsequent growth increment. The direction of propagation at each tip node is determined by the principle of local symmetry, which asserts that a crack will extend in a direction that locally eliminates the mode-II loading as \cite{cotterell1980slightly}
\begin{equation}
\frac{\theta_g}{2}=\arctan \big( \frac{1}{4}\left[\frac{K_I}{K_{I I}}-\operatorname{Sgn}\left(K_{I I}\right) \sqrt{\left(\frac{K_I}{K_{I I}}\right)^2+8}\right] \big)
\end{equation}
where $\operatorname{Sgn}\left(K_{I I}\right)$ denotes the sign of $K_{I I}$.
The magnitude of this advancement, $\Delta a$, is regulated by a bilinear growth law that relates the propagation distance to an effective mode-I stress intensity factor $\overline{K}_I$ as
\begin{equation}
\frac{\Delta a}{\Delta t}=\left\{\begin{array}{cc}0 &, \hspace{0.5cm}\bar{K}_I<K_{I c} \\ \dot{a}_o\left(\frac{\bar{K}_I-K_{I c}}{K_o}\right)^m &, \hspace{0.5cm}\bar{K}_I>K_{I c}\end{array}\right.
\end{equation}
where $\bar{K}_I$ is the mode-I stress intensity factor defined as
\begin{equation}
\bar{K}_I=K_I \cos ^3\left(\frac{\theta_g}{2}\right)-\frac{3}{2} K_{I I} \cos \left(\frac{\theta_g}{2}\right) \sin \theta_g
\end{equation}
and $\dot{a}_o$, $K_o$ and $m$ are model constants. The simulator calculates a target advance step size as a fraction of a characteristic element length near the tip. This ensures that the growth increment is proportional to the local mesh resolution, preventing excessive element stretching in a single step. The algorithm evaluates this growth law to determine the scalar advance distance and direction for each tip node. This information is then used to compute a three-dimensional propagation vector, which is added to the coordinates of the frontal nodes of the specialized crack tip elements, thereby physically extending the fracture domain.

\subsubsection{Adaptive Element Management and Mesh Quality Control}

Simply advancing the crack tip nodes leads to rapid degradation of mesh quality, as the elements immediately behind the front would become unacceptably elongated. To counteract this, a continuous and adaptive process of element splitting, merging, and geometric correction is performed to maintain a well-conditioned mesh. This logic is governed by a set of heuristic criteria based on element geometry and aspect ratios.

The process is controlled by a characteristic length $l_c$, which is dynamically calculated as the averaged dimension of all active crack tip elements. This characteristic length is used to define a target size $d_{smax}$ for the standard 8-node elements that are adjacent to the 9-node crack-tip elements. The remeshing algorithm iterates through these adjacent elements, evaluating the lengths of their sides, i.e. $d_{sl}$ and $d_{sr}$, relative to the crack tip element.

A primary concern is element stretching. The algorithm follows four distinct cases for managing element refinement based on these lengths (see Figure \ref{fig:remesh1}):
\begin{itemize}
\item Case 1 (Tip Element Splitting):  A quality check on the aspect ratio $d_l/d_w$ of the 9-node crack tip elements is performed. If the aspect ratio $d_l / d_w$ exceeds a prescribed tolerance, the single tip element is subdivided into two new smaller tip elements. This is a complex operation that also forces the splitting of the adjacent quadrilateral element into a new configuration of elements to maintain mesh conformity.

\item Case 2 (Adjacent Quadratic Element Split): If both the left and right sides of the adjacent element become stretched such that $d_{sl} > d_{smax}$ and $d_{sr} > d_{smax}$, the algorithm determines that the adjacent element is too large. It then splits this 8-node element by introducing new nodes and forming a new 8-node quadrilateral element between the old element and the crack tip element, as shown in Figure \ref{fig:remesh2}.
\item Case 3 (Adjacent Triangular Element Split): If only one side of the element, for example the left side $d_{sl}$, exceeds the target length $d_{smax}$, the distortion is localized. In this scenario, a new 6-node triangular element is formed on that side to manage the transition, while the other side remains unchanged.
\item Case 4 (No Split): If neither side exceeds the target length, no new elements are formed, and the existing element is simply stretched to accommodate the small advancement of the crack tip.
\end{itemize}
This operation introduces new nodes and elements, increasing the global element count, and updates the nodal connectivity arrays to reflect the new topology.

\begin{figure}[h!]
    \centering
    \includegraphics[width=0.48\textwidth]{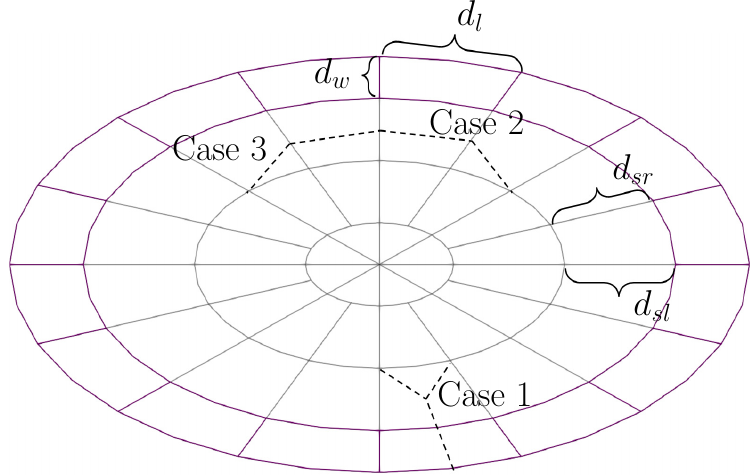}
    \caption{Schematic of remeshing strategy concerning element stretching.}
    \label{fig:remesh1}
\end{figure}

Conversely, as the fracture front moves forward, regions that were previously near the tip become part of the fracture's interior. In these areas, a high mesh density is no longer necessary and can be computationally wasteful. The algorithm identifies opportunities for mesh coarsening by detecting specific element configurations. When a 6-node triangular element exists adjacent to an 8-node quadrilateral element, a geometric quality check is performed. This check calculates the dot product of the vectors forming the angle at the shared node and compares it to a critical tolerance, e.g., $\cos(45^\circ)$. If the angle is found to be too obtuse, indicating the triangle is becoming unacceptably flat and distorted, the algorithm flags it. In a subsequent step, the 6-node triangle and the adjacent 8-node quadrilateral are merged back into a single, larger, and more well-conditioned 8-node quadrilateral element. This dual process of refinement and coarsening maintains a computationally efficient mesh throughout the simulation, concentrating resolution at the propagating front where solution gradients are highest. The implementation of the adaptive remesh in details is summarized in Algorithm \ref{alg:remeshing}.

\subsubsection{Solution Field Projection}

After the mesh topology and nodal coordinates have been updated, the physical solution fields, including relative crack face displacements, fluid pressure, and the nodal age for the leak-off calculation, would be transferred from the old mesh to the new one. This projection is essential for providing an accurate initial guess for the subsequent Newton-Raphson iteration and for preserving the history of time-dependent processes.

To avoid expensive global mesh destructions and memory fragmentation, the projection logic operates on a highly localized state-machine paradigm. Prior to tip advancement, the engine pre-allocates static one-dimensional and two-dimensional floating-point snapshot buffers in the stack memory, creating a copy of the active nodes' transient displacement, pressure, and time fields. Following the localized coordinate modifications, an inverse mapping algorithm triangulates the parametric spatial weights of the new topology within the host elements of the previously stored snapshot. The standard $\mathbf{U}$ and $p$ solution arrays are subsequently faithfully interpolated. 

A critical mathematical singularity is introduced for newly birthed nodes deposited exactly at the moving crack tip, as their absolute coordinates now lie entirely outside the geometrical domain of the old mesh. Without intervention, a naive interpolation extrapolation could trigger severe unphysical fluid volume expansions at the fracture extremum. To strictly enforce continuity across time steps, the algorithm mathematically regularizes these out-of-bounds nodes by applying a heuristic displacement reduction constraint scaled relative to surrounding local element dimensions. Simultaneously, the temporal exposure age $t_{\text{node}}$ of these nascent tip nodes is reset to the current global simulation time, ensuring the Carter leak-off integration evaluates accurately against the freshly created fracture surface. This localized buffer-projection scheme strictly isolates the complicated sparse array pointer assignments from the broader $\mathcal{O}(N_{el})$ global mesh, enabling agile, distortion-free fracture front evolution.

\begin{figure}
	\renewcommand{\figurename}{Algorithm} 
	\renewcommand{\thefigure}{1}
	
	\hrule height 0.8pt 
	\vspace{2pt}
	\hrule height 0.4pt
	\vspace{5pt}
	
	\caption{Adaptive Remeshing Procedure}
	\label{alg:remeshing}
	
	\addtocounter{figure}{-1}
	
	\begin{algorithmic}[1]
		\Procedure{AdaptiveRemesh}{$t^i, \mathcal{M}^i, \mathbf{U}^i, \mathbf{P}^i$}
		
		\Statex \hspace{-0.1cm} Stage 1: Fracture Front Advancement
		\For{each crack tip node $j$ in mesh $\mathcal{M}^i$}
		\State Compute stress intensity factors $K_I(j)$ and $K_{II}(j)$
		\Statex \hspace{\algorithmicindent} from solution $\mathbf{U}^i$.
		\State Determine advance vector $\Delta\mathbf{x}(j)$ from LEFM
		\Statex \hspace{\algorithmicindent} growth law.
		\State Update tip node: $\mathbf{x}_j^{\text{new}} \gets \mathbf{x}_j^i + \Delta\mathbf{x}(j)$.
		\EndFor
		
		\Statex \hspace{-0.1cm} Stage 2: Adaptive Element Management
		\State Initialize new mesh $\mathcal{M}^{\text{new}}$ with advanced tip nodes.
		\For{each element $e$ in the remeshing zone}
		\State Evaluate element size and aspect ratio.
		\If{element length exceeds target size $d_{\text{max}}$}
		\State Split element $e$ into smaller elements in $\mathcal{M}^{\text{new}}$.
		\ElsIf{geometric criteria for coarsening are met}
		\State Merge element $e$ with neighbors in $\mathcal{M}^{\text{new}}$.
		\EndIf
		\EndFor
		\State Finalize new mesh topology $\mathcal{M}^{i+1}$.
		
		\Statex \hspace{-0.1cm} Stage 3: Solution Field Projection
		\For{each new or moved node $k$ in $\mathcal{M}^{i+1}$}
		\State Find old element $e_{\text{old}}$ containing the new
		\Statex \hspace{\algorithmicindent} location $\mathbf{x}_k^{i+1}$.
		\State Compute local coordinates $\boldsymbol{\eta}_k$ within $e_{\text{old}}$.
		\State Project solution fields (Displacement $\mathbf{U}$,
		\Statex \hspace{\algorithmicindent} Pressure $P$) via interpolation using shape
		\Statex \hspace{\algorithmicindent} functions of $e_{\text{old}}$.
		\If{node $k$ is a new crack tip}
		\State Scale projected displacement $\mathbf{U}_k^{i+1}$ for
		\Statex \hspace{\algorithmicindent} volume conservation.
		\State Set nodal age $t_{\text{node},k} \gets t^{i+1}$ for leak-off
		\Statex \hspace{\algorithmicindent} calculation.
		\EndIf
		\EndFor
		
		\Return Updated mesh and solution state
		\Statex \hspace{\algorithmicindent} $(\mathcal{M}^{i+1}, \mathbf{U}^{i+1}, \mathbf{P}^{i+1})$
		\EndProcedure
	\end{algorithmic}
	
	\vspace{5pt}
	\hrule height 0.4pt
\end{figure}

\section{Computational Performance: Optimization and Parallelism}\label{sec:para}

The practical utility of the advanced simulation platform is contingent not only on the physical fidelity of its models but also on its computational performance. For the HyFrac.fun platform, the Symmetric Galerkin Boundary Element Method (SGBEM) at the core of its numerical engines presents significant performance challenges. The primary computational bottleneck is the formation and assembly of the global stiffness matrix for the solid mechanics $\mathbf{A}$, which arises from the discretization of the weak-form traction integral equation (Eqs.~\ref{eq:fracbie} and \ref{eq:wellbie}). The calculation of this dense symmetric matrix involves a double surface integral over all pairs of elements on the fracture surface. For a mesh with $N_{el}$ elements, the computational complexity of forming $\mathbf{A}$ from scratch is $\mathcal{O}(N_{el}^2)$. As fractures grow, the number of elements increases, and this quadratic complexity leads to prohibitive runtimes, rendering large-scale long-duration simulations intractable.

The optimization strategies presented in this section were originally conceived by Mood (2019)~\cite{mood2019coupled} for a dimension-reduced SGBEM formulation for height-contained fracture in which fracture geometry is represented as a one-dimensional line mesh. In the present work, these strategies are adapted and applied to a fully three-dimensional non-planar SGBEM formulation, where the fracture surface is an arbitrarily curved two-dimensional surface mesh and the stiffness operator $\mathcal{A}$ involves double surface integrals rather than line integrals. The fully 3D geometry makes the mesh topology changes at every time step significantly more complex than the height-contained case. Additionally, the dynamic solver-switching algorithm (Section~\ref{sec:solvers_combined}) and the adaptive remeshing algorithm (Section~\ref{sec:remeshing}) are documented in full detail here.

\subsection{Incremental Matrix Updates and Memory Layout Optimization}

\begin{figure}[h!]
    \centering
    \includegraphics[width=0.4\textwidth]{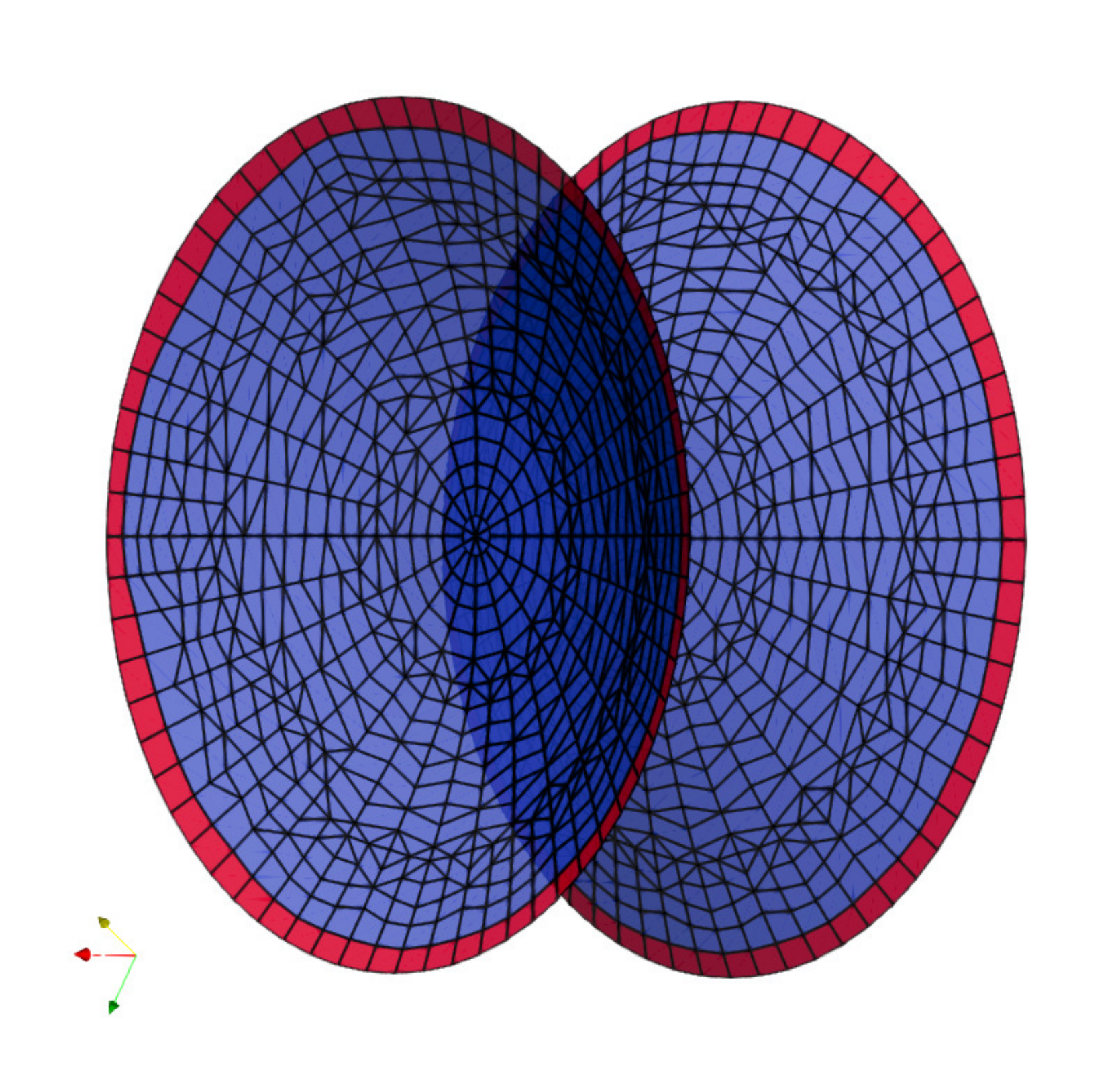}
    \caption{Schematic of tip element (red) and stationary elements (blue).}
    \label{fig:remesh2}
\end{figure}

A key physical observation during a hydraulic fracturing simulation is that fracture propagation, and therefore the associated remeshing, is a highly localized phenomenon. Changes to the mesh geometry are confined to small zones at the advancing crack tips. The vast majority of elements, those constituting the interior of the fracture, remain geometrically stationary from one time step to the next (see Figure \ref{fig:remesh2}). Since the entries of the stiffness matrix $\mathbf{A}$ are solely a function of the geometry and relative position of element pairs, the large portion of the matrix corresponding to interactions between pairs of stationary elements is invariant between time steps. Recomputing these entries at every step is computationally wasteful.

\begin{figure}[htbp]
  \centering 
  \includegraphics[width=0.21\textwidth]{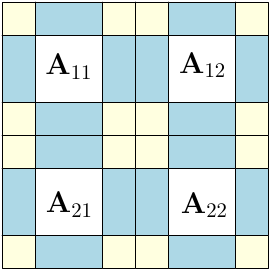}
  \hfill 
  \includegraphics[width=0.21\textwidth]{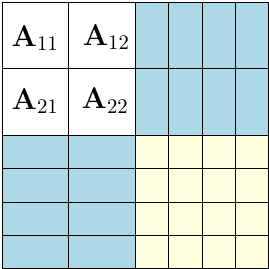}

  \caption{Permutation of the stiffness matrix $\mathbf{A}$ via element renumbering. (Left) A natural ordering of elements results in scattered update blocks (blue for evolved-stationary interaction and yellow for updated-updated interaction). (Right) The re-ordered matrix, which groups all stationary elements (white), concentrating the update blocks into a contiguous region.}
  \label{fig:simple_pair}
\end{figure}

To exploit this temporal locality, a data caching or memoization strategy is implemented. The full stiffness matrix $\mathbf{A}$ is computed only once at the beginning of the simulation or after a major remeshing event that adds many new stationary elements. The contributions from all pairs of stationary elements are stored in a persistent, auxiliary matrix, denoted \texttt{gk\_temp} in the Fortran sources.

At the beginning of each subsequent time step, within the main simulation loop of the simulation routine, the current stiffness matrix \texttt{gk} is initialized by copying the pre-computed values from \texttt{gk\_temp}. The algorithm then proceeds to compute only those entries that are affected by the recent crack growth. This is managed through a status flag, \texttt{iflag\_elem\_update(i)}, for each element \texttt{i}. The expensive double integral calculation for an element pair `(i, j)` is guarded by a conditional check:
\begin{lstlisting}[language=Fortran]
if ((iflag_elem_update(i) /= 0).or.(iflag_elem_update(j) /= 0)) then
    ! ... perform expensive double integral calculation ...
endif
\end{lstlisting}
This logic ensures that computations are only performed if at least one of the elements in the pair belongs to a remeshed zone. This incremental update strategy reduces the computational complexity of the matrix formation at each time step from $\mathcal{O}(N_{el}^2)$ to a more favorable $\mathcal{O}(M \cdot N_{el})$, where $M$ is the number of elements in the small, remeshed tip zones ($M \ll N_{el}$).

To seamlessly map this mathematical incremental scheme into hardware-efficient operations, the platform executes a dynamic topological reorganization. This logical separation is physically implemented by renumbering the element indices prior to matrix assembly. As illustrated in Figure~\ref{fig:simple_pair}, a natural element numbering (left panel) scatters the active degrees of freedom randomly throughout the global arrays. Updating the affected matrix coordinates using such an ordering enforces highly strided non-continuous memory access patterns. Because modern multi-core processors rely heavily on hierarchical L1/L2 cache systems, these scattered look-ups cause catastrophic cache-line misses and severe latency penalties.

To circumvent this, the engine generates an active permutation topology array that strictly clusters all geometrically stationary elements (i.e. white blocks in Figure~\ref{fig:simple_pair}) first, immediately followed by the elements within the dynamically evolving fracture-tip zones. Upon applying this permutation, the matrix segments requiring heavy integrational updates are forcibly concentrated into a single contiguous block residing in the bottom-right corner of the global matrix (see right panel of Figure~\ref{fig:simple_pair}). Consequently, when the processor fetches a 1D packed-array cache line, it can continuously process localized mathematical operations, dramatically accelerating the computation of the dense $\mathbf{A}$ matrix.

The benefit of this permutation is most evident when considering the memory layout. Since the matrix is symmetric, only the lower (or upper) triangle is stored in a packed one-dimensional array. In Fortran's column-major layout, the re-ordered update region corresponds to a few dense contiguous vertical stripes at the end of the packed array. This memory layout is highly cache-friendly. When a processor core fetches a cache line to compute one matrix entry, the subsequent entries it needs tend to be in the same or an adjacent cache line, drastically reducing memory latency and improving the overall computational throughput.

\subsection{Shared-Memory Parallelization and Load Balancing Strategy}

To accelerate the computationally intensive matrix formation and assembly routines, the simulation engine is parallelized for shared-memory multi-core processors using OpenMP. The implementation targets the primary computational loops that iterate over element pairs for the $\mathcal{O}(N_{el}^2)$ SGBEM matrix and over individual elements for the $\mathcal{O}(N_{el})$ FEM matrices. The core of this strategy involves distributing the expensive element-level calculations across all available threads, followed by a synchronized assembly into the global system matrices.

The parallel region in the main $\mathcal{O}(N_{el}^2)$ stiffness routine, which leverages the incremental update logic, is defined as follows:

\begin{lstlisting}[language=Fortran]
!$omp parallel default(shared) private(...)
    !$omp do schedule(dynamic) 
    do ii=1,nelem
        i = mordered_elems(ii)
        ! ... outer loop body ...
        do jj=1,ii
            j = mordered_elems(jj)
            ! Conditional check for incremental update
            if ((iflag_elem_update(i) /= 0) .or. (iflag_elem_update(j) /= 0)) then
                ! ... expensive element-pair integral calculation ...
                
                ! Synchronized assembly for main stiffness matrix
!$omp critical (assmgkd)
                call assmgkadd(ek,no,ni,nodes(1,i),nodes(1,j))
!$omp end critical (assmgkd)

                ! Synchronized assembly for saved stiffness matrix
                if (iflag_elem_update(i) == 2 .and. ...) then
!$omp critical (assmgkd1)
                    call assmgkadd1(ek,no,ni,nodes(1,i),nodes(1,j))
!$omp end critical (assmgkd1)
                endif
            endif
        end do
    end do
!$omp end do
!$omp end parallel
\end{lstlisting}

A critical challenge in this parallel assembly is the avoidance of race conditions, where multiple threads might attempt to write to the same memory location in the global stiffness matrix simultaneously. This implementation handles the challenge by enforcing mutual exclusion using \texttt{! \$OMP CRITICAL} directives around all global assembly calls (e.g., \texttt{assmgkadd} and \texttt{assmgkadd1}) . This construct ensures that only one thread at a time can execute the assembly code block, guaranteeing data integrity at the cost of serializing this portion of the workload.

To maximize processor utilization under this model, a hybrid load-balancing strategy is adopted using different OpenMP scheduling clauses:

\begin{enumerate}
    \item Dynamic Scheduling for Non-Uniform Workloads: For the $\mathcal{O}(N_{el}^2)$ SGBEM matrix assembly, the nested loop structure (\texttt{do jj=1,ii}) creates a highly non-uniform, triangular workload. Iterations for elements with a high index \texttt{ii} perform substantially more work than those with a low index. For this loop, the \texttt{schedule(dynamic)} clause is employed. This allows threads that finish a short iteration to dynamically request a new one from the runtime pool, thus balancing the load and preventing thread idleness.
    
    \item Static Scheduling for Uniform Workloads: For the $\mathcal{O}(N_{el})$ assembly of sparse FEM matrices (e.g., in \texttt{cross\_stiff\_fluid} for generating coupling blocks $\mathbf{B}$ and $\mathbf{B}^T$ and global force vectors (e.g., in \texttt{formgf\_only} for generating global load vectors $\mathbf{R}_q$, $\mathbf{R}_p$, $\mathbf{R}_c$, and $\mathbf{R}_f$), the computational work per element is relatively constant. In these routines, the \texttt{schedule(static)} clause is used. This approach assigns iterations to threads in fixed, pre-determined chunks, which minimizes runtime scheduling overhead and is highly efficient for such well-balanced workloads.
\end{enumerate}

This combined strategy of algorithmic optimization (incremental updates), memory-layout optimization (matrix permutation), and a hybrid parallel scheduling model (dynamic and static) is essential to the performance of the simulation engine. It transforms the computationally bound $\mathcal{O}(N_{el}^2)$ problem into a parallel algorithm that scales effectively on modern multi-core processors, making the on-demand, interactive simulation model a practical reality.

\subsection{Advanced Linear Solvers and Computational Stability for the Coupled System}
\label{sec:solvers_combined}

The monolithic solution of the fully coupled, linearized system of algebraic equations (Eqs.~\ref{eq:compact_solid},\ref{eq:compact_fluid},\ref{eq:compact_darcy},\ref{eq:compact_channel}) is a critical step demanding robust and specialized linear solvers due to the complex, heterogeneous nature of the Jacobian matrix. This matrix comprises a dense, symmetric solid mechanics block $\mathbf{A}$, a sparse fluid flow block $\mathbf{C}$, and non-zero coupling blocks $\mathbf{B}$ and $\mathbf{B}^T$. The solution strategy is hierarchical, employing both iterative methods optimized for performance and a dynamic direct solver fallback to guarantee stability across a wide range of non-linear conditions.

The simulation strategically deploys two variants of Preconditioned Krylov subspace methods, with the choice dictated by the symmetry properties of the coupled matrix imposed by the fluid model. When the fluid model allows the linearized system to be treated as symmetric (e.g., during the Picard iteration or for initial linear estimates), the Mixed Preconditioned Conjugate Gradient (PCG) method is utilized. Conversely, for non-Newtonian power-law fluids solved via the Newton-Raphson scheme, the resulting Jacobian may be non-symmetric. In these instances, the specialized Mixed Stabilized Bi-Conjugate Gradient (BiCGSTAB) method is automatically activated to reliably handle the potentially non-symmetric system.

To accelerate the convergence of both iterative solvers, a Block-Jacobi Preconditioner is consistently employed. This preconditioner exploits the block structure of the Jacobian by utilizing the inverse of the diagonal blocks of the solid and fluid components (i.e. matrices $\mathbf{A}$ and  $\mathbf{C}$ respectively), offering a performance advantage over a simpler diagonal preconditioner.

In massively concurrent simulation environments hosted on the cloud, achieving absolute fault tolerance is paramount; a solitary numerical oscillation artifact cannot be allowed to irrevocably crash a live user rendering session. Numerical fragility consistently manifests during extreme spatial events especially when intense stress-shadow interference forces adjacent propagating fractures to violently buckle and warp out-of-plane. Under these pathological conditions, the mesh topology often experiences extreme localized stretching before a successful remeshing operation can catch it, generating a Jacobian matrix defined by severe asymmetry and a highly ill-conditioned eigenvalue spread. When standard accelerated Krylov methods operate on such degraded preconditioners, the iterative residual norm frequently oscillates or succumbs to non-linear divergence.

To structurally insulate the system from these chaotic bounds, an autonomic degradation and fault-recovery closed loop assumes control of the execution stream (see Algorithm \ref{alg:solver_switch}). Whenever the engine detects that an iterative convergence trajectory has exceeded critical thresholds or residual bounds are aggressively breached, it immediately interrupts the deteriorating cycle. The module triggers a hot-swap fallback into the robust LAPACK direct matrix factorization scheme (\texttt{dspsv}), accepting the temporary computational slowdown in exchange for mathematical certitude while navigating the singularity. Once the engine registers a continuous progression of completely successful direct iterations, which demonstrates that the broader matrix condition number has geometrically healed following subsequent adaptive remeshing, the control stream aggressively reverts the linear protocol back into the high-performance Preconditioned Conjugate Gradient iterative mode.

\begin{figure}
	\renewcommand{\figurename}{Algorithm}
	\renewcommand{\thefigure}{2}
	
	\hrule height 0.8pt 
	\vspace{2pt}
	\hrule height 0.4pt
	\vspace{5pt}
	
	\caption{Dynamic Solver Switching Procedure}
	\label{alg:solver_switch}
	
	\addtocounter{figure}{-1}
	
	\begin{algorithmic}[1]
		\Procedure{SolveCoupledSystem}{$\mathbf{J}, \mathbf{R}, \mathbf{U}^{k}, \text{Mode}_{\text{prev}}$}
		\State $\text{Mode}_{\text{current}} \gets \text{Mode}_{\text{prev}}$
		\State $\text{FailureCount} \gets 0$
		\State $\text{MaxFailures} \gets 5$ \Comment{Threshold for sustained failure}
		\State $\text{RevertSteps} \gets 20$ \Comment{Steps required to revert from Direct mode}
		
		\Loop
		\If{$\text{Mode}_{\text{current}} = \text{DIRECT}$}
		\State $\mathbf{U}^{k+1}, \text{ierr} \gets \text{SolveDirect}(\mathbf{J}, \mathbf{R})$
		\Else
		\State $\text{Solver} \gets \text{ChooseIterative}(\mathbf{J}.\text{Symmetry})$ \Comment{PCG or BiCGSTAB}
		\State $\mathbf{U}^{k+1}, \text{ierr} \gets \text{SolveIterative}(\mathbf{J}, \mathbf{R}, \text{Solver})$
		\EndIf
		
		\If{$\text{ierr} \neq 0$} \Comment{Failure Detected}
		\State $\text{Mode}_{\text{current}} \gets \text{DIRECT}$ \Comment{Switch to robust mode}
		\State $\text{FailureCount} \gets \text{FailureCount} + 1$
		\If{$\text{FailureCount} > \text{MaxFailures}$} 
		\State error: Unstable
		\EndIf
		\Else \Comment{Success}
		\State $\text{FailureCount} \gets 0$
		\If{$\text{Mode}_{\text{current}} = \text{DIRECT}$ and $\text{ConsecutiveSuccess} \ge \text{RevertSteps}$}
		\State $\text{Mode}_{\text{current}} \gets \text{ITERATIVE}$ \Comment{Return to high-performance mode}
		\EndIf
		\EndIf
		
		\Return $\mathbf{U}^{k+1}, \text{Mode}_{\text{current}}$
		\EndLoop
		\EndProcedure
	\end{algorithmic}
	
	\vspace{5pt}
	\hrule height 0.4pt
\end{figure}

\subsection{Parallel Scalability and Performance Analysis}
\begin{figure}[htbp]
	\centering
	\begin{subfigure}[b]{1.0\linewidth}
		\centering
		\includegraphics[width=\linewidth]{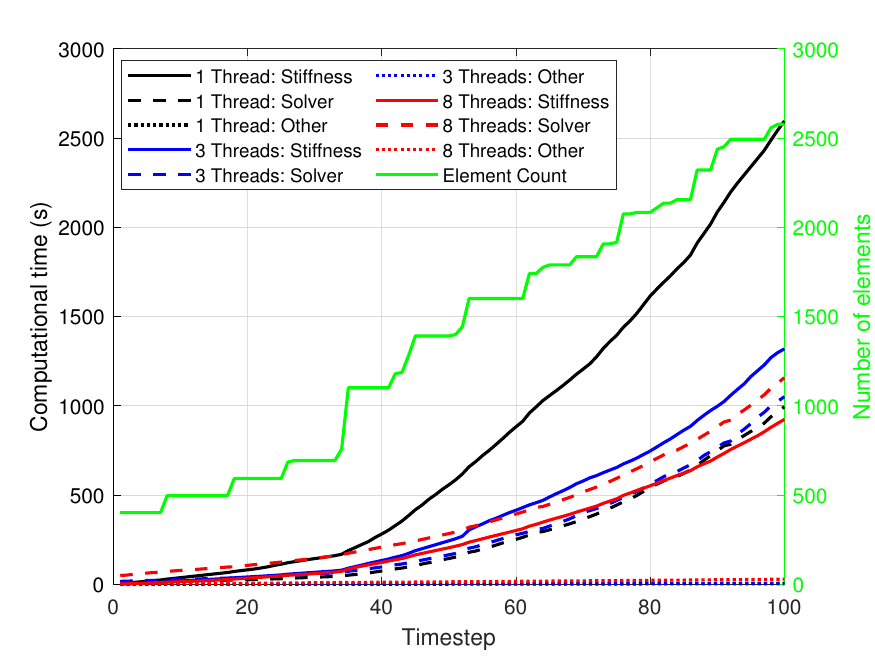}
		\caption{}
		\label{fig:sub_time}
	\end{subfigure}
	
	\vspace{0.5em} 
	
	\begin{subfigure}[b]{1.0\linewidth}
		\centering
		\includegraphics[width=\linewidth]{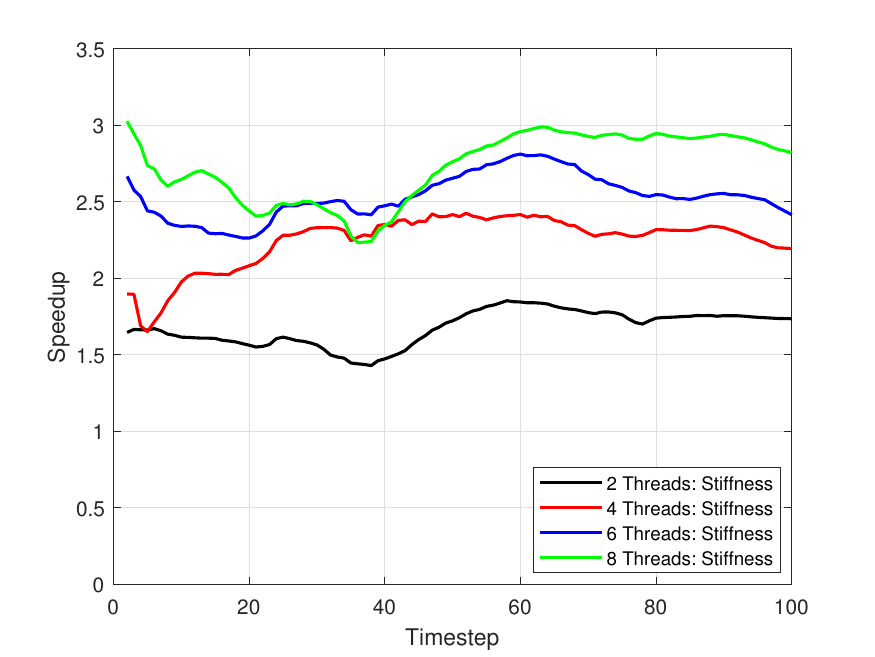}
		\caption{}
		\label{fig:sub_speedup}
	\end{subfigure}
	
	\vspace{0.5em} 
	
	\begin{subfigure}[b]{1.0\linewidth}
		\centering
		\includegraphics[width=\linewidth]{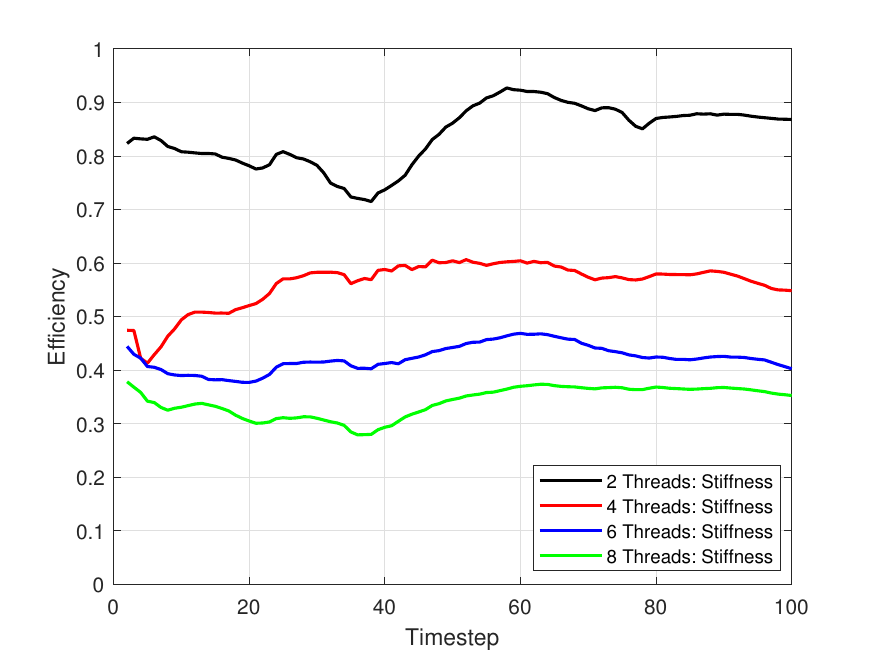}
		\caption{}
		\label{fig:sub_efficiency}
	\end{subfigure}
	
	\caption{Performance analysis of the shared-memory parallelization strategy. (a) Breakdown of computational time versus simulation time steps for varying thread counts. (b) Scaling speedup of the stiffness assembly routine. (c) Parallel efficiency (Speedup/Threads).}
	\label{fig:performance_analysis}
\end{figure}

To validate the effectiveness of the parallelization strategies, a performance benchmark was conducted on a representative hydraulic fracturing scenario involving continuous crack growth over 100 time steps. The results are summarized in Fig. \ref{fig:performance_analysis}.

Fig. \ref{fig:performance_analysis}(a) presents the breakdown of computational time by major code component. The green line tracks the number of elements $N_{el}$, which grows almost linearly with time steps. 
The single-threaded stiffness assembly time (black solid line) exhibits a distinct non-linear growth that correlates strongly with the element count, empirically confirming the theoretical $O(N_{el}^2)$ complexity of the SGBEM formulation. By increasing the thread count to 8 (red solid line), the absolute time required for stiffness assembly is drastically reduced. While the linear solver and other routines do not benefit from parallelization, the stiffness assembly remains the dominant cost, justifying the focus on optimizing this specific routine.

The scalability of the implementation is further quantified in Fig. \ref{fig:performance_analysis}(b) and (c). The speedup is calculated relative to the single-threaded execution time. As shown in Fig. \ref{fig:performance_analysis}(b), the implementation achieves a speedup of approximately three times using 8 threads. The curves exhibit fluctuations corresponding to remeshing events where the matrix size changes abruptly. Fig. \ref{fig:performance_analysis}(c) reveals that parallel efficiency decreases from around $0.8$ at 2 threads to around $0.35$ at 8 threads. This decay is attributed to two factors: (1) the serialization overhead introduced by the \texttt{OMP CRITICAL} sections required for thread-safe matrix assembly, and (2) memory bandwidth saturation, as multiple cores compete to write to the large shared global stiffness matrix. Despite the sub-linear scaling, the absolute reduction in runtime shown in Fig. \ref{fig:performance_analysis}(a) is critical for enabling interactive, cloud-based simulations.

To further enhance the scalability of the cloud platform, particularly for massive fracture networks where memory contention is acute, a more sophisticated domain decomposition strategy is required. Implementing an advanced graph partitioning mechanism that introduces logical cuts within the stiffness matrix stripes would allow for finer-grained load balancing across arbitrary thread counts, significantly mitigating the synchronization overhead. Furthermore, as the matrix assembly is optimized, the linear solver increasingly becomes the dominant computational bottleneck. Future development will therefore focus on replacing the current hybrid solver with fully parallelized preconditioned Krylov subspace algorithms to ensure that the solution phase scales commensurately with the assembly phase.

\section{Cloud System Software Architecture Design and Implementation}
\label{sec:cloud_arch}

The numerical engines and performance optimizations described in Sections~\ref{sec:formulations} and~\ref{sec:para} are necessary but not sufficient for the platform's core contribution: the automated lifecycle simulation from fracture propagation to well production. The engineering realization of the automated mesh-and-state handoff between the two engines, which is enabled by the operator isomorphism established in Section~\ref{sec:unified_operators}, requires a cloud-native orchestration layer capable of managing the full simulation lifecycle as a single, reproducible, server-side workflow. This section describes the specifically designed orchestration layer. Traditional numerical simulation software relies heavily on thick clients, necessitating complex environment configurations, localized HPC hardware, and specialized visualization tools. To break these technical barriers, the cloud-native simulation platform HyFrac.fun utilizes a three-tier service-oriented architecture (SOA), ensuring the deep decoupling of the presentation layer, the application logic layer, and the data rendering layer.

\begin{figure*}[htbp]
	\centering
	\includegraphics[width=1.0\textwidth]{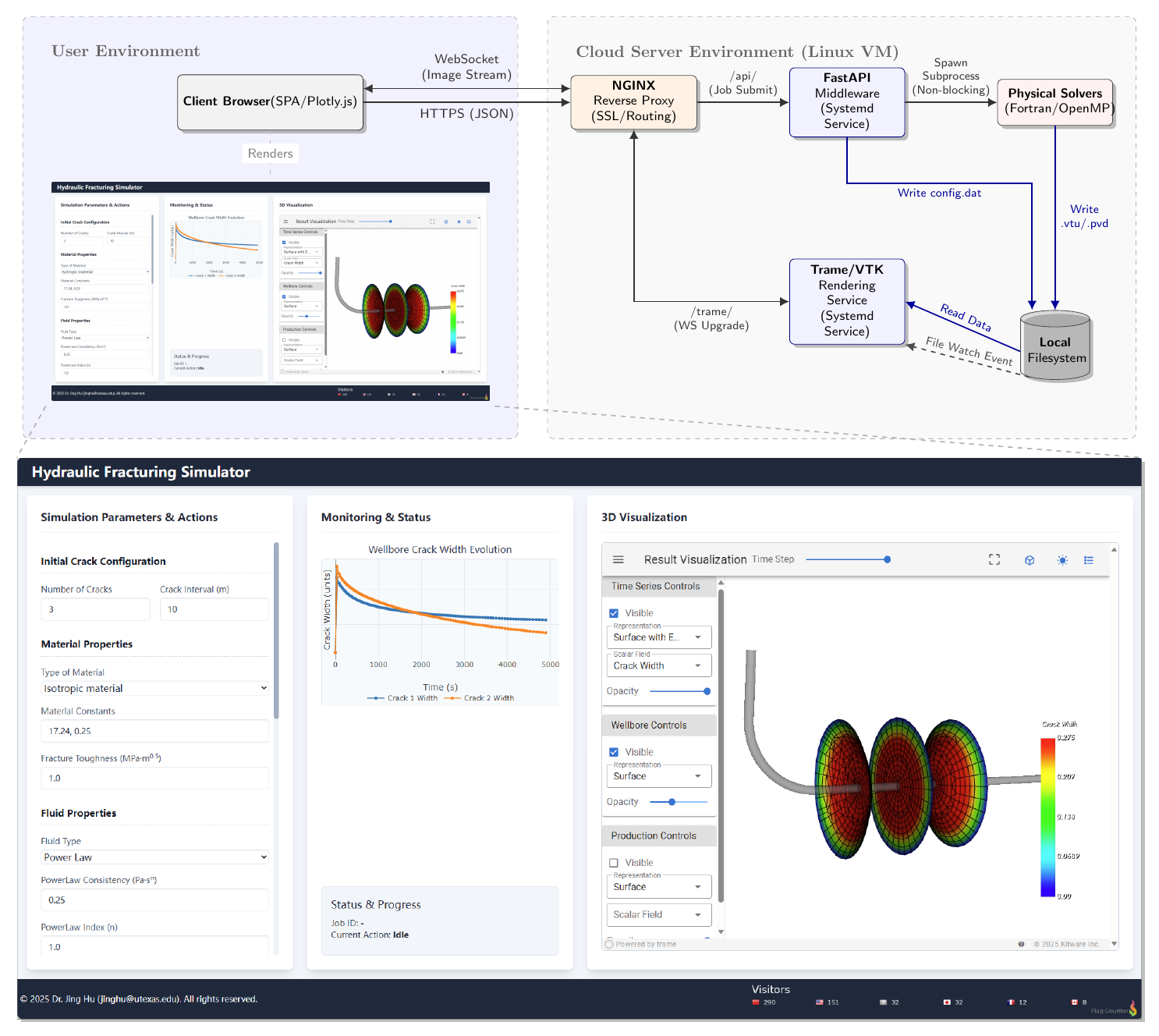}
	\caption{Overview of the HyFrac.fun cloud platform. The architecture links the responsive user interface (above left) to the legacy Fortran solvers (above right) via an Nginx gateway, employing FastAPI for process management and Trame for server-side rendering.}
	\label{fig:architecture}
\end{figure*}

As shown in Fig. \ref{fig:architecture}, the systemic architecture safely isolates the Fortran computational engines within a sandbox backend, while users engage with the physics simulation entirely through a standard responsive web browser. The three microservice layers, i.e. presentation layer, application logic layer, and data processing and rendering layer, communicate via strictly defined Application Programming Interfaces (APIs) and full-duplex WebSockets.

\subsubsection{Presentation Layer: Responsive Front-End}
The presentation layer is deployed as an ultra-lightweight Single Page Application (SPA), utilizing HTML5, CSS3, ES6+ JavaScript, and the Tailwind CSS component framework. The layout is visually segmented into a parameter input sidebar, a real-time tracking dashboard, and an embedded 3D spatial viewport (see Fig. \ref{fig:architecture}).

Complex interactions are strictly controlled. When engineering parameters such as cluster spacing or fluid rheology properties are validated and submitted, the native \texttt{Fetch API} serializes the payload into a JSON structure, asynchronously dispatching it to the backend. Subsequently, the client instantiates a non-blocking asynchronous polling timer that routinely requests job timeline state-machine fragments every 3 seconds, thereby minimizing bandwidth consumption while providing immediate workflow feedback.

\subsubsection{Application Logic Layer: Asynchronous Backend Middleware}
The core orchestration of the platform heavily features a FastAPI-based asynchronous middleware. This acts as the crucial bridge connecting the stateless HTTP protocol to the legacy Fortran solver core execution. Leveraging the Asynchronous Server Gateway Interface (ASGI) and Pydantic object models, it prevents server deadlocks effectively.


The state machine tracking the simulation lifecycle ensures robust transitions from 'Queued' to 'Running' to 'Completed'. Because parallel high-fidelity mechanical simulations can quickly dominate filesystem I/O, the middleware executes a strict anti-conflict isolation locking mechanism:
\begin{enumerate}
    \item Process Polling and Zombie Eradication: Utilizing system-level commands (e.g., \texttt{ps -o pid,stat}), the FastAPI middleware continually probes the active operation tables. If a solver instance is maliciously deadlocked or structurally hanging, it forces a \texttt{SIGKILL} cleanup before deploying the queued computation, freeing the directory sandbox lock and resuming operations seamlessly without blocking the HTTP web loop.
    \item RegEx I/O Sniffer: Because the primary Fortran binary executes as a closed \texttt{subprocess.Popen} entity, conventional API progression bars fail. To circumvent this, the middleware initializes an aggressive regular expression File System Sniffer to scan for generated VTU output strings (\texttt{simulation\_output\_t*.vtu}) on the storage interface. The maximal numerical suffix dynamically correlates to the internal timestep, retroactively computing a progress percentage asynchronously to send directly back the client.
\end{enumerate}

\subsubsection{Data Processing and Rendering Layer: Cloud-Streamed Visualization}
\begin{figure*}[htbp]
	\centering
	\begin{tikzpicture}[
		font=\sffamily\small,
		node distance=0.6cm,
		basebox/.style={rectangle, draw=black!80, thick, rounded corners=2pt, align=center, fill=white, drop shadow={opacity=0.2, shadow xshift=2pt, shadow yshift=-2pt}},
		storage/.style={basebox, fill=blue!10, minimum width=8cm},
		daemon/.style={basebox, fill=gray!5, dashed, inner sep=15pt},
		component/.style={basebox, fill=white, minimum height=0.7cm, minimum width=2cm},
		vtkstep/.style={basebox, fill=orange!5, font=\sffamily\scriptsize, minimum width=1.5cm},
		gateway/.style={basebox, fill=green!10, minimum width=8cm},
		client/.style={basebox, fill=yellow!10, minimum width=3.8cm},
		arrow/.style={-{Stealth[scale=1.0]}, thick, draw=black!70},
		labeltext/.style={font=\sffamily\tiny\itshape, color=black!70}
		]
		
		\node[storage] (data) {
			\textbf{Cloud Sandbox File System} \\ 
			\texttt{*.vtu / *.pvd} (Unstructured Grid Sequence)
		};
		
		\node[daemon, below=1.2cm of data, minimum width=14cm, minimum height=4.2cm] (daemon_bg) {};
		\node[anchor=north west, font=\sffamily\bfseries\footnotesize] at (daemon_bg.north west) {Data Processing \& Rendering Layer (Python Daemon)};
		
		\node[component, below=0.4cm of daemon_bg.north, xshift=-4cm] (watcher) {\textbf{Watchdog} \\ FS Monitor};
		\node[component, below=0.4cm of daemon_bg.north, xshift=4cm] (trame) {\textbf{Trame} \\ State Controller};
		
		\node[vtkstep, below=1.8cm of daemon_bg.north, xshift=-5.2cm] (v1) {vtkXMLReader \\ (Parser)};
		\node[vtkstep, right=0.6cm of v1] (v2) {vtkDataSetMapper \\ (Scalar LUT)};
		\node[vtkstep, right=0.6cm of v2] (v3) {vtkActor \\ (Materials)};
		\node[vtkstep, right=0.6cm of v3] (v4) {vtkRenderer \\ (Camera/Lights)};
		\node[vtkstep, right=0.6cm of v4] (v5) {vtkRenderWindow \\ (Off-screen)};
		
		\begin{scope}[on background layer]
			\node[fill=orange!10, draw=orange!30, fit=(v1) (v5), inner sep=8pt, label={[font=\tiny\bfseries, orange]north:VTK 3D Rendering Pipeline}] (vtk_group) {};
		\end{scope}
		
		\node[gateway, below=3.5cm of daemon_bg.north] (nginx) {
			\textbf{Network Ingress (Nginx)} \\ 
			WebSocket Protocol Upgrade \& SSL Proxy
		};
		
		\node[client, below=0.8cm of nginx, xshift=-2.1cm] (ui) {\textbf{Frontend UI (Vue.js)} \\ Parameter Panels / Charts};
		\node[client, below=0.8cm of nginx, xshift=2.1cm] (canvas) {\textbf{Remote View} \\ HTML5 Canvas};
		
		\draw[arrow, dashed] (data.south) -- ++(0,-0.4) -| (watcher.north) node[midway, above, labeltext, xshift=-2cm, yshift=-0.2cm] {FS Events};
		\draw[arrow] (data.south) -- ++(0,-0.4) -| (v1.north) node[midway, left, labeltext, yshift=0.3cm] {Binary Stream};
		
		\draw[arrow] (watcher) -- (trame) node[midway, above, xshift=-0.3cm, labeltext] {Trigger Reload};
		\draw[arrow] (trame) -| (v2.north) node[pos=0.2, above, xshift=0.3cm, labeltext] {LUT Update};
		\draw[arrow] (trame) -| (v4.north) node[pos=0.2, above, xshift=-0.2cm, labeltext] {Pose Sync};
		
		\draw[arrow] (v1) -- (v2);
		\draw[arrow] (v2) -- (v3);
		\draw[arrow] (v3) -- (v4);
		\draw[arrow] (v4) -- (v5);
		
		\draw[arrow] (v5.south) -- ++(0,-0.3) -| ($(nginx.north)+(1cm,0)$) node[midway, right, xshift=0.3cm, yshift=-0.2cm, labeltext] {Video Stream};
		\draw[arrow, <->] (trame.south) -- ++(0,-0.3) -| ($(nginx.north)+(-1cm,0)$) node[midway, left,xshift=0.1cm, labeltext] {WSS State Binding};
		
		\draw[arrow, <->] ($(nginx.south)+(-2.1cm,0)$) -- (ui.north) node[midway, left, labeltext] {User Events};
		\draw[arrow] ($(nginx.south)+(2.1cm,0)$) -- (canvas.north) node[midway, right, labeltext] {Image Frames};
		
	\end{tikzpicture}
	\caption{Layered architecture of the cloud-native rendering system. The diagram illustrates the top-down flow from binary storage through the VTK server-side rendering pipeline to the interactive web client via a WebSocket-enabled Nginx gateway.}
	\label{fig:rendering_architecture}
\end{figure*}
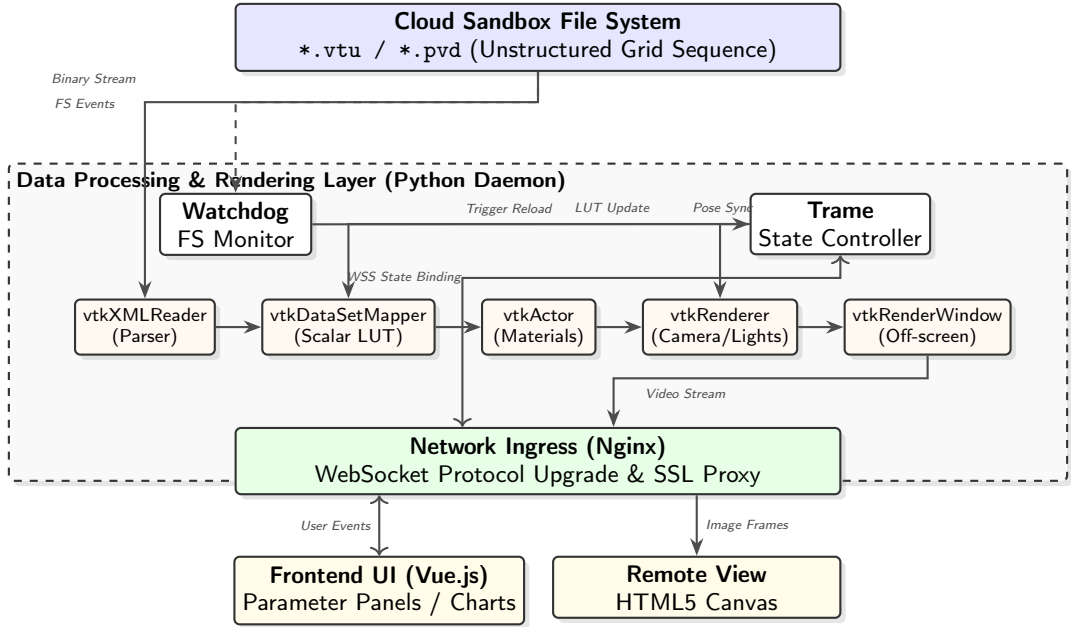

Broadcasting gigabyte-scale, highly transient, non-structured 3D grids over standard internet connections immediately exhausts client-side graphic allocation limits. To solve this, a remote server-side rendering pipeline utilizing the VTK (Visualization Toolkit) and Trame framework is embedded inside the cloud node (Fig \ref{fig:rendering_architecture}).

A specialized Python daemon acts as the dedicated scene server. \texttt{vtkXMLUnstructuredGridReader} interfaces natively mount the output matrices from disk, bridging them through the \texttt{vtkDataSetMapper}. Control variables (e.g., the camera azimuth, pressure contour bounds, and slice transparency) are bound to UI components bidirectionally via WebSockets. Interactions made on the client browser push microscopic parameter pulses upstream. The backend scene graph updates instantly, re-rasterizes the graphic frame onto the remote GPU, and pushes a high-definition, encoded compressed media stream directly into the local Canvas hook. A Python \texttt{watchdog} continuously guards the file system during a simulation; as new timestep dumps populate, the script hot-reloads the VTK pipeline, dynamically progressing the geometry exactly mirroring the underlying physical time domain without human intervention.

\subsubsection{Cloud-Native Infrastructure and Application Routing}
To orchestrate global routing streams, handle static asset caching, and establish an encrypted SSL/TLS perimeter across these independent microservice nodes, the platform employs an Nginx reverse-proxy ingress controller.

\begin{lstlisting}[language=bash, basicstyle=\ttfamily\small, caption={Nginx Reverse-Proxy Gateway and WebSocket HTTP-Upgrade Configuration}, label={lst:nginx_config}]
server {
	listen 443 ssl http2;
	server_name www.hyfrac.fun;
	
	# 1. Static asset routing for frontend SPA
	location /app/ {
		alias /var/www/hyfrac_frontend/dist/;
		try_files  / /index.html;
	}
	
	# 2. Transactional routing for FastAPI backend
	location /api/ {
		proxy_pass http://127.0.0.1:8000;
		proxy_set_header Host System.Management.Automation.Internal.Host.InternalHost;
		proxy_set_header X-Real-IP ;
		proxy_set_header X-Forwarded-For ;
	}
	
	# 3. Streaming and WebSocket upgrade for rendering
	location /trame/ {
		proxy_pass http://127.0.0.1:8080;
		proxy_http_version 1.1;
		# Capture and execute protocol upgrade handshake
		proxy_set_header Upgrade ;      
		proxy_set_header Connection "upgrade";       
		# Increase timeout to prevent disconnection
		proxy_read_timeout 86400;                    
	}
}
\end{lstlisting}

By executing a protocol upgrade from standard HTTP towards persistent WebSockets in the \texttt{/trame/} routing block (Code Snippet \ref{lst:nginx_config}), the system successfully blends extreme high-performance offline calculations with ultra-low latency interactivity, validating the migration of traditional mechanics codebases onto containerized Web architectures.

\section{Numerical Verification and Examples}
\label{sec:examples}

To demonstrate the capabilities of the HyFrac.fun platform across the full hydraulic fracturing lifecycle, several benchmark simulations are performed. The goal is to showcase the simulator's ability to capture complex three-dimensional fracture interaction phenomena during propagation and to subsequently analyze the production performance of the resulting geometries. While the results will be visualized and analyzed through the HyFrac.fun web interface, this section outlines the setup for each case and discusses the expected qualitative and quantitative outcomes based on prior research.

\subsection{Verification of the Numerical Engines}
\label{subsec:verification}

To ensure the reliability of the cloud-orchestrated numerical architecture, a series of benchmark verifications are performed. These cases are designed to confirm that the coupled SGBEM-FEM framework correctly captures the mechanical response of the rock and the steady-state fluid transport within the porous medium. In both cases, the hybrid solver is tested in the limit of negligible fluid viscosity to isolate and verify the fundamental integral formulations against classical analytical solutions.

The accuracy of the SGBEM elasticity solver is first evaluated against the analytical solution for a circular penny-shaped crack of radius $R$ embedded in an infinite, isotropic, linear elastic medium. Under a uniform internal pressure $p$, the analytical opening profile $w(r)$ as a function of the radial distance $r$ from the center is given by Sneddon \cite{Sneddon1946}:
\begin{equation}
	w(r) = \frac{8(1-\nu^2)pR}{\pi E} \sqrt{1 - \left(\frac{r}{R}\right)^2}
	\label{eq:sneddon}
\end{equation}
where $E$ denotes Young’s modulus and $\nu$ represents Poisson’s ratio. Within the HyFrac.fun platform, this condition is simulated by injecting fluid at a constant rate. To approximate the constant pressure distribution required by Eq.~(\ref{eq:sneddon}), a negligible fluid viscosity ($\mu = 1 \times 10^{-6}$~Pa$\cdot$s) is prescribed, minimizing the pressure gradient within the fracture channel. In the simulation, we adopted a Young's modulus of $E = 17.24$~GPa, a Poisson's ratio of $\nu = 0.25$, the final constant pressure of $p = 279$ MPa and the crack radius of $R = 10.028$~m.

\begin{figure*}[htbp]
	\centering
	\begin{subfigure}[b]{0.48\textwidth}
		\centering
		\includegraphics[width=\textwidth]{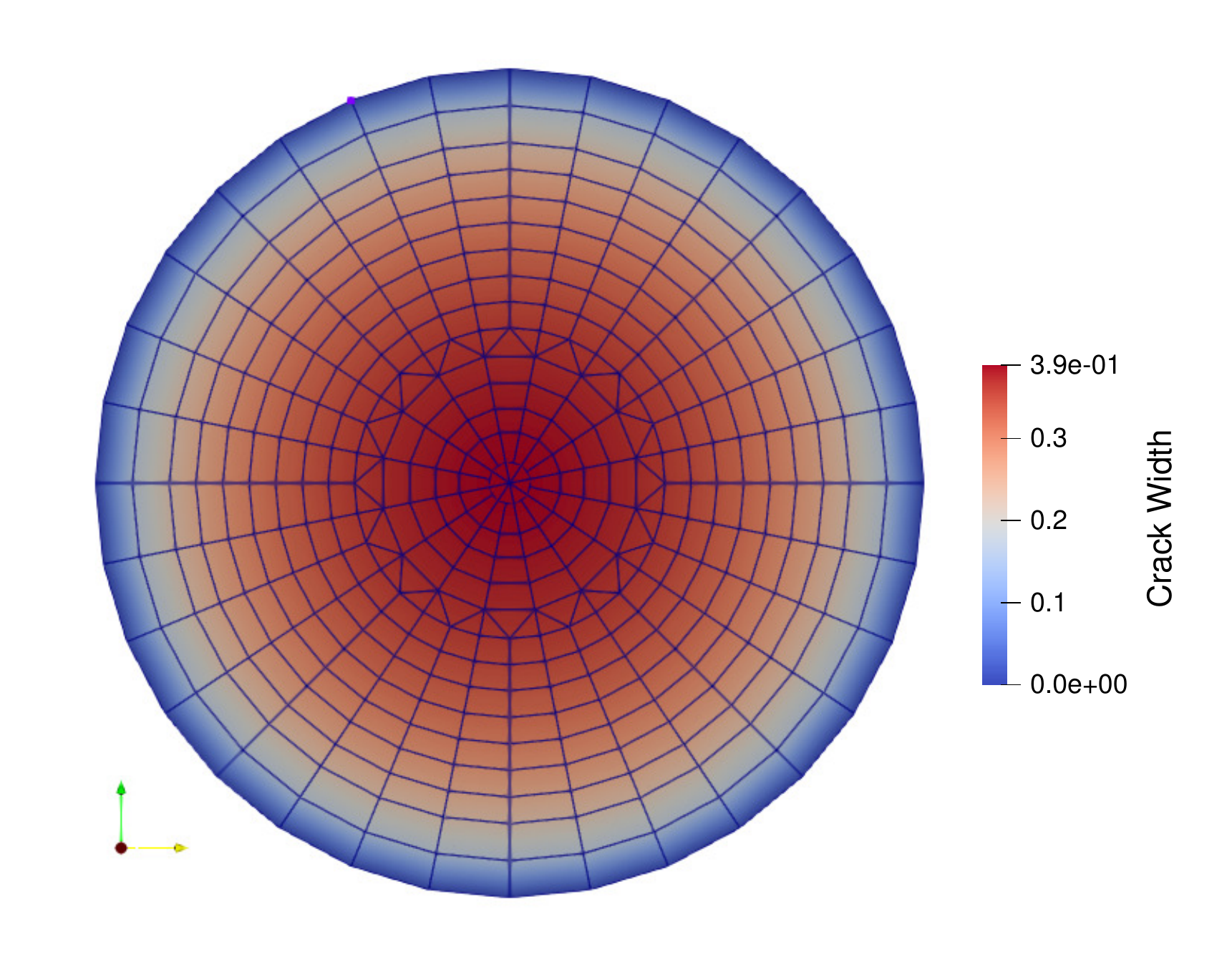}
		\caption{Distribution of simulated fracture width $w(r)$.}
		\label{fig:widnum}
	\end{subfigure}
	\hfill
	\begin{subfigure}[b]{0.48\textwidth}
		\centering
		\includegraphics[width=\textwidth]{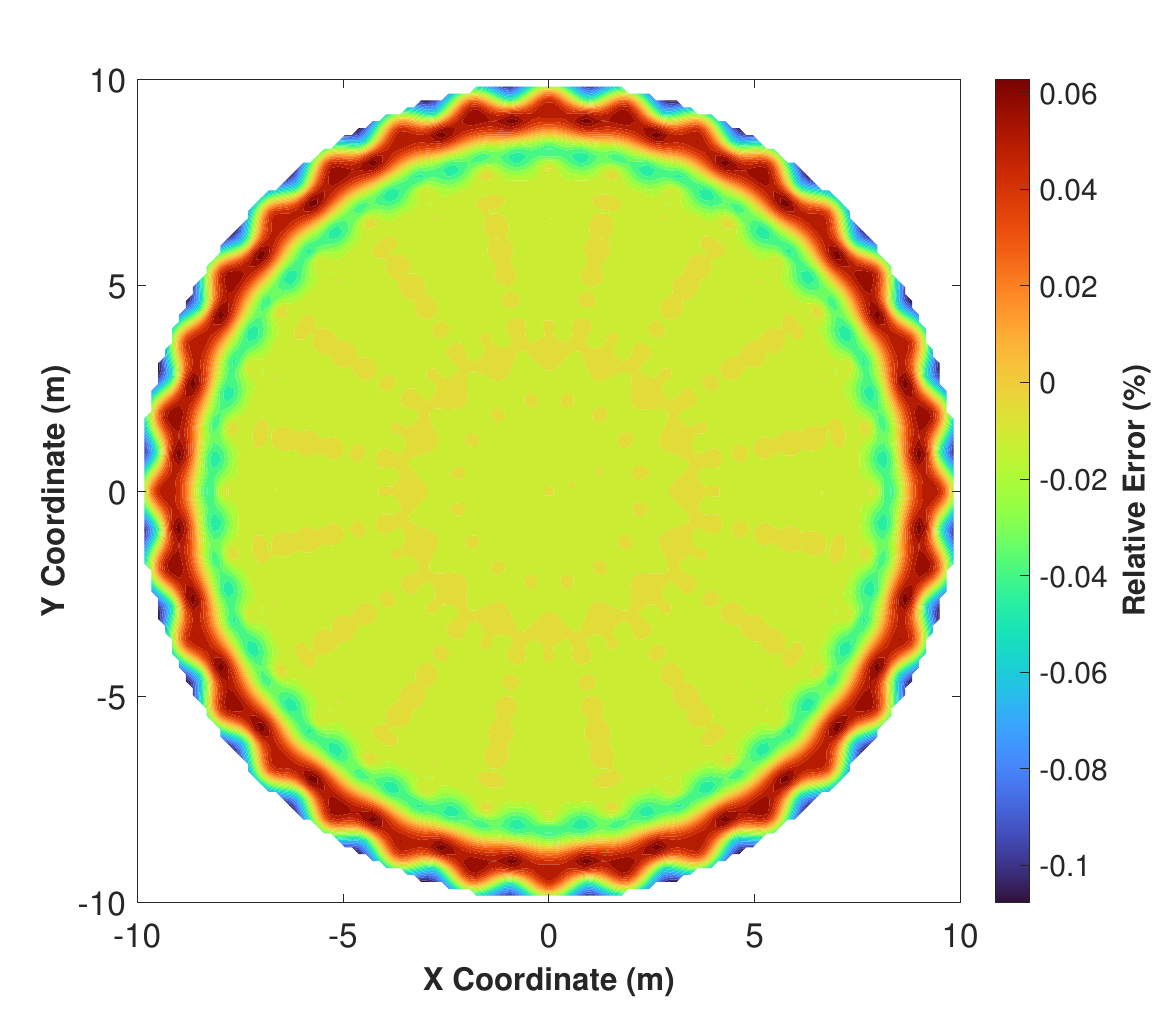}
		\caption{Relative error (\%) of fracture width.}
		\label{fig:widval}
	\end{subfigure}
	
	\vspace{0.5cm} 
	
	\begin{subfigure}[b]{0.48\textwidth}
		\centering
		\includegraphics[width=\textwidth]{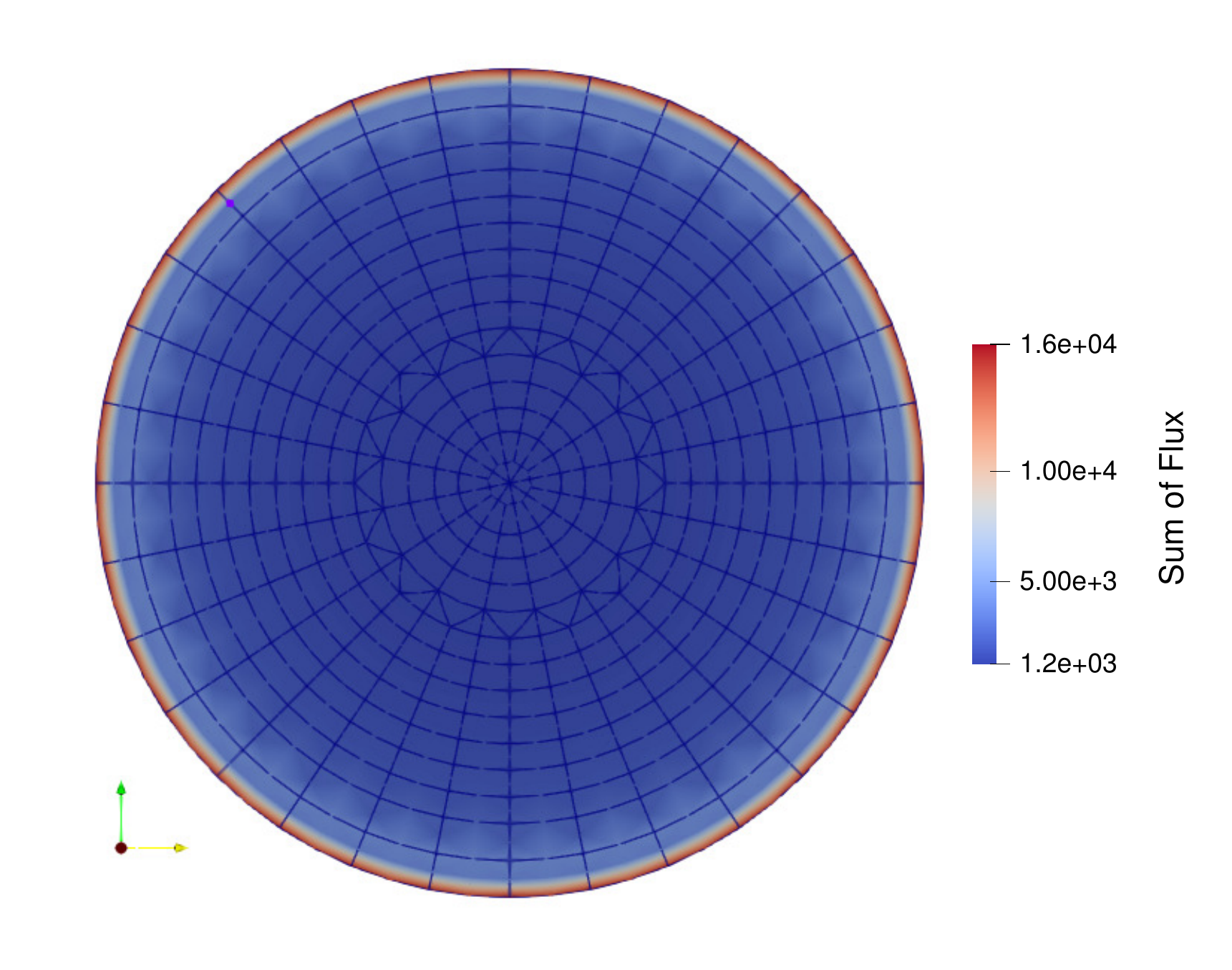}
		\caption{Distribution of the sum of flux $\Sigma q$.}
		\label{fig:fluxnum}
	\end{subfigure}
	\hfill
	\begin{subfigure}[b]{0.48\textwidth}
		\centering
		\includegraphics[width=\textwidth]{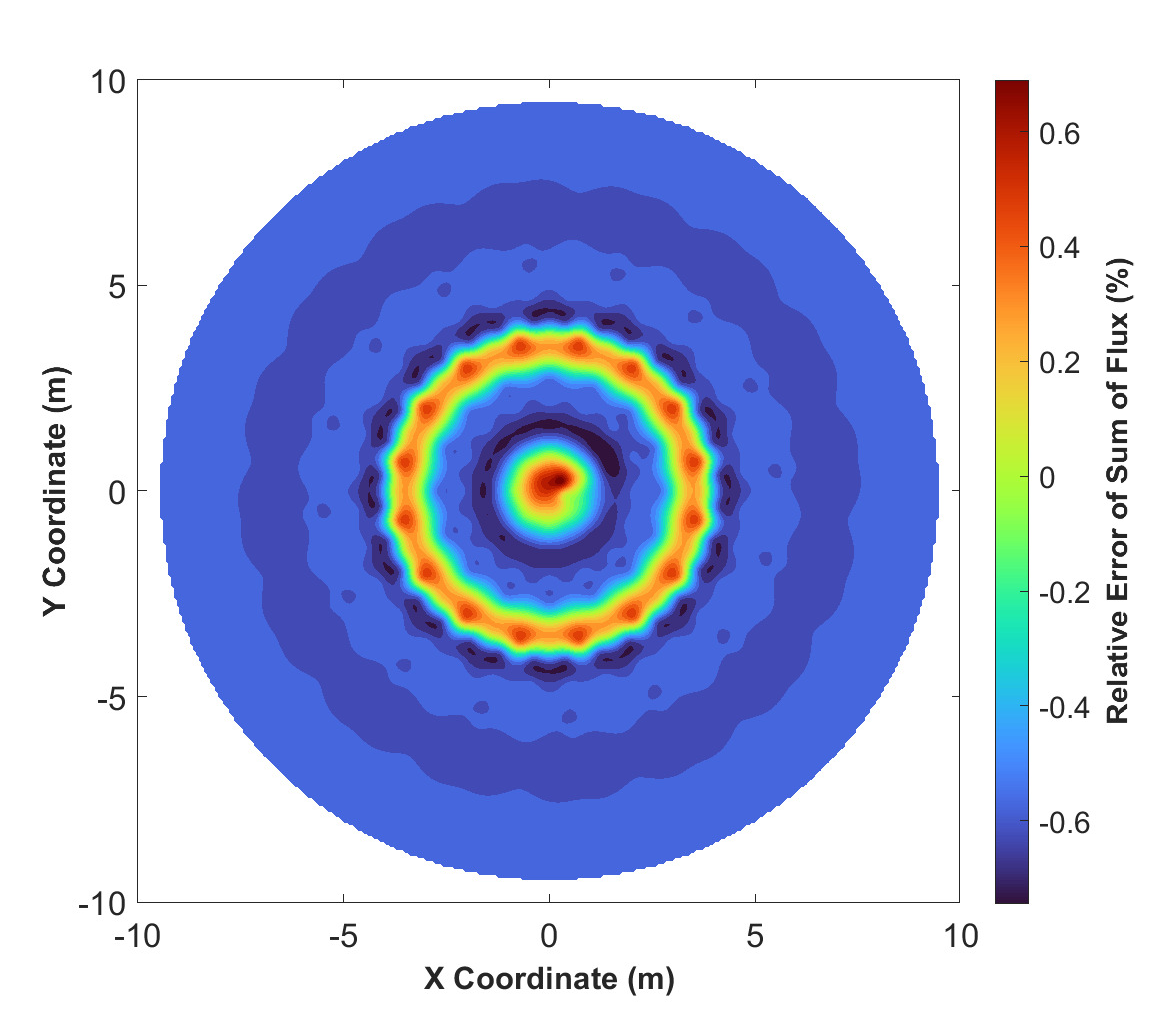}
		\caption{Relative error (\%) of sum of flux.}
		\label{fig:fluxval}
	\end{subfigure}
	
	\caption{Numerical verification of the HyFrac.fun platform after 100 simulation timesteps. The top row (a--b) validates the mechanical elasticity solver against Sneddon's solution. The bottom row (c--d) validates the reservoir-fracture coupling against the analytical Darcy flux distribution.}
	\label{fig:verification_grid}
\end{figure*}

The numerical distribution of the fracture width after 100 simulation timesteps is presented in Fig.~\ref{fig:widnum}, displaying the characteristic elliptical profile. To quantify the precision of the solver, the relative error between the numerical results and the analytical Sneddon solution is plotted in Fig.~\ref{fig:widval}. The error remains consistently low across the fracture surface, confirming the high fidelity of the SGBEM formulation and its ability to maintain numerical stability over extended simulation durations.

Following the mechanical validation, the steady-state production engine is verified by evaluating the coupling between Darcy flow in the reservoir and channel flow within the fracture. This verification follows the benchmark established by Hu and Mear (2022) \cite{Hu2022} for a stationary penny-shaped fracture in a porous matrix. For a circular fracture with constant internal pressure, the distribution of the sum of flux $\Sigma q$, which represents the total fluid exchange between the reservoir and both fracture faces, is expressed analytically as:
\begin{equation}
	\Sigma q = -\frac{\kappa}{\mu_{res}} \frac{4 (p_{\infty} - p_{f})}{R \pi \sqrt{1-\left(r / R\right)^2}}
	\label{eq:darcy_flux_dist}
\end{equation}
where $\kappa$ is the permeability of the porous medium, and $(p_{\infty} - p_{f})$ represents the pressure drawdown.  In the simulation, we adopted a permeability of $\kappa = 100$~Darcy, reservoir fluid viscosity $\mu_{res}=1$~cP, and  pressure drawdow of $(p_{\infty} - p_{f}) = -100$ GPa.

Consistent with the mechanical verification, the same low-viscosity limit is utilized to ensure a uniform pressure $p_f$ across the fracture surface. The numerical distribution of the sum of flux after 100 timesteps is shown in Fig.~\ref{fig:fluxnum}. Figure~\ref{fig:fluxval} illustrates the relative error between the numerical result and the analytical solution. Note that the physical quantities on the crack front are th intensity of singularity of the singular sum of flux, and the singular sum of flux in the region very close to the tip is not shown in Figure~\ref{fig:fluxnum}. The small relative error demonstrates that the cloud-based platform preserves the precision of core numerical research while delivering results through a modern service-oriented architecture.

\subsection{Multi-Stage Fracture Interaction and Production Lifecycle}
\label{subsec:complex_interaction}

The secondary phase of the numerical analysis demonstrates the platform's capability to model the complete lifecycle of a complex hydraulic fracturing treatment. In unconventional reservoir development, the ultimate goal of numerical simulation is not merely to predict the final fracture geometry, but to evaluate how that specific geometry dictates long-term hydrocarbon recovery. This involves simulating the non-planar propagation of multiple interacting fractures and seamlessly transitioning those finalized 3D meshes into a steady-state production analysis. To investigate these coupled behaviors, a configuration of three parallel fractures initiated simultaneously from a horizontal wellbore is examined.

The baseline elastic properties of the formation are set with a Young's modulus of $E = 17.24$~GPa, a Poisson's ratio of $\nu = 0.25$, and a fracture toughness of $K_{IC} = 1.0$~MPa$\sqrt{\text{m}}$. Fluid is injected into each fracture at an equal and constant rate of $0.01$~m$^3$/s, assuming zero leak-off, for a total of 130 simulation timesteps. To systematically evaluate the system, three distinct cases are configured by varying the initial crack interval (cluster spacing) and the fracturing fluid rheology. 

\subsubsection{Effect of Cluster Spacing on Stress Shadowing}

The proximity of simultaneously propagating fractures induces a severe mechanical interference known as the stress shadow effect. As a fracture opens, it compresses the surrounding rock matrix, significantly altering the local stress for adjacent fractures.  To quantify this phenomenon, we compare Case 1, featuring a narrow crack interval of $3$~m, against Case 2, which utilizes a wider interval of $6$~m. Both cases employ a low-viscosity Newtonian fracturing fluid ($\mu = 0.01$~Pa$\cdot$s).

\begin{figure*}[htbp]
	\centering
	\begin{subfigure}[b]{0.32\textwidth}
		\centering
		\includegraphics[width=\textwidth]{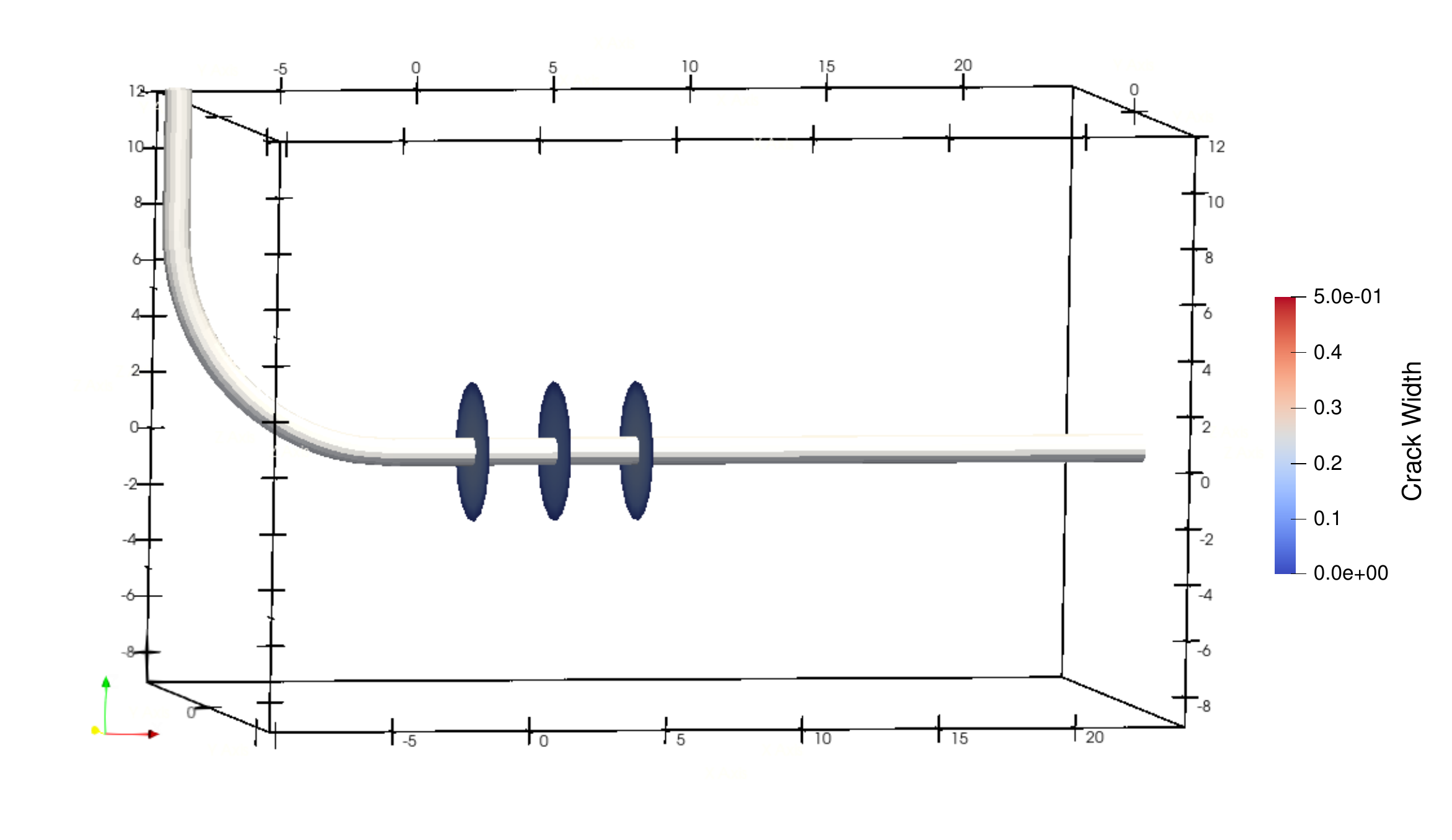}
		\caption{Case 1: Step 1}
		\label{fig:c1_s1}
	\end{subfigure}
	\hfill
	\begin{subfigure}[b]{0.32\textwidth}
		\centering
		\includegraphics[width=\textwidth]{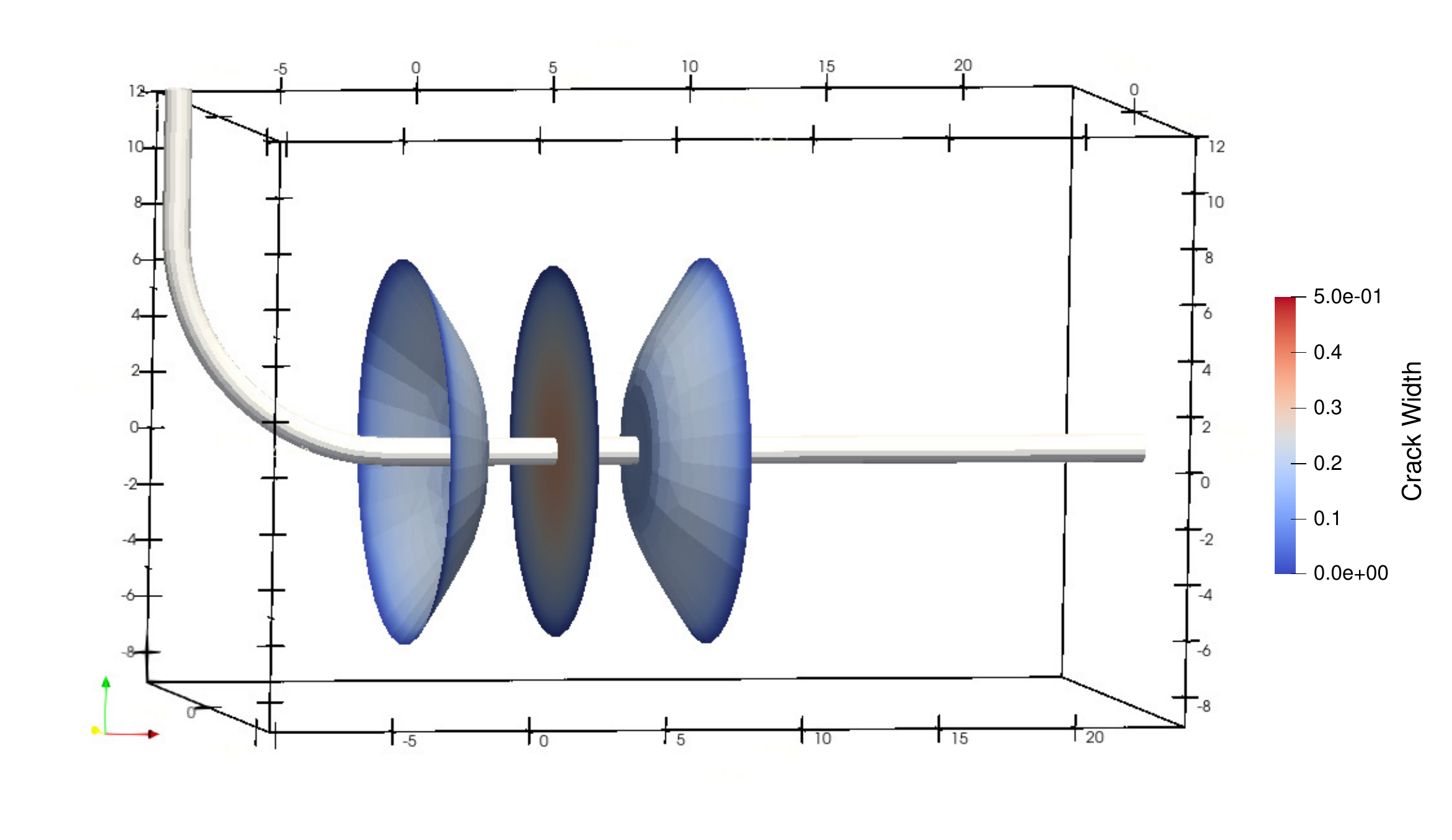}
		\caption{Case 1: Step 65}
		\label{fig:c1_s65}
	\end{subfigure}
	\hfill
	\begin{subfigure}[b]{0.32\textwidth}
		\centering
		\includegraphics[width=\textwidth]{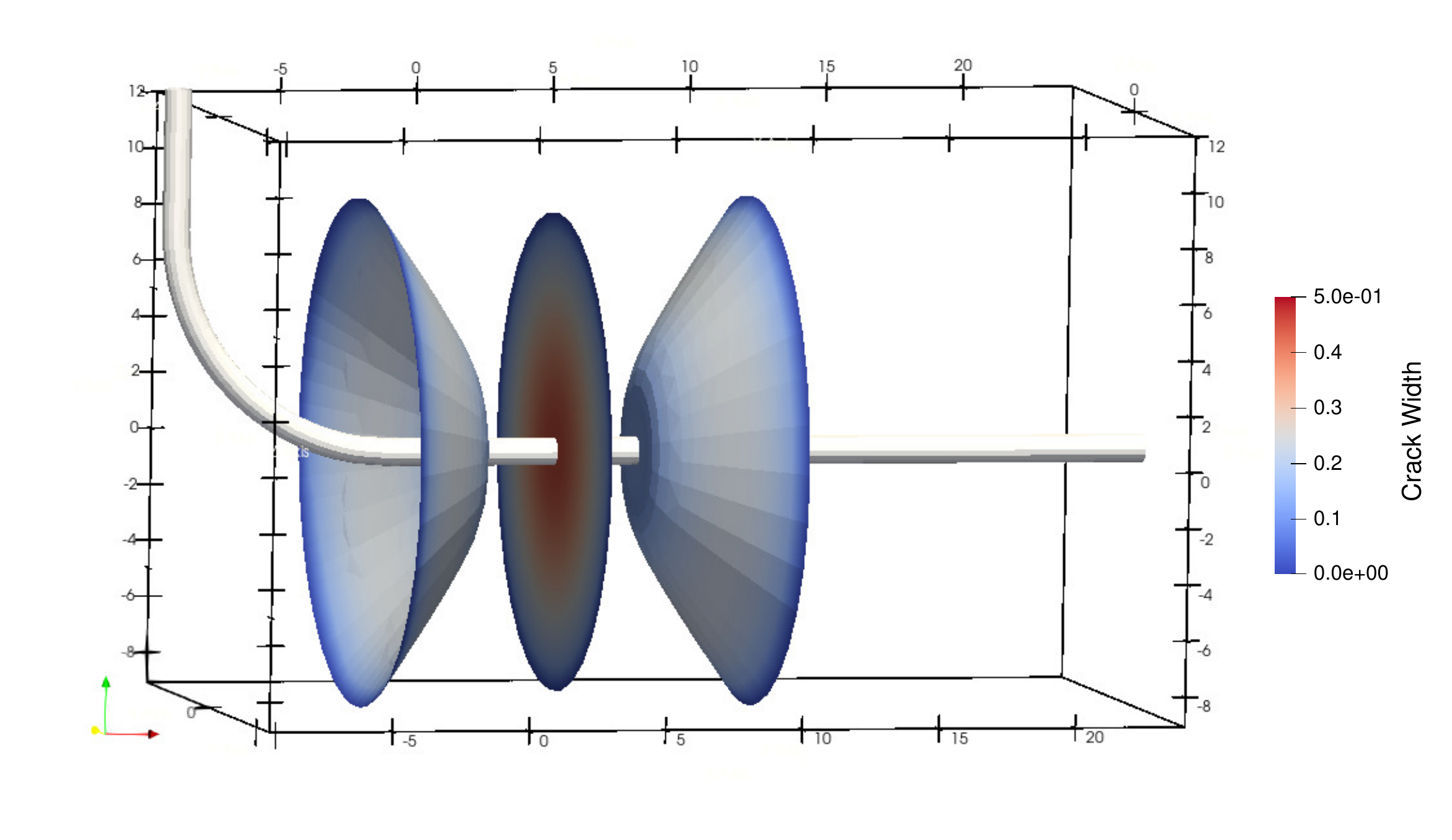}
		\caption{Case 1: Step 130}
		\label{fig:c1_s130}
	\end{subfigure}
	
	\vspace{0.3cm}
	
	\begin{subfigure}[b]{0.32\textwidth}
		\centering
		\includegraphics[width=\textwidth]{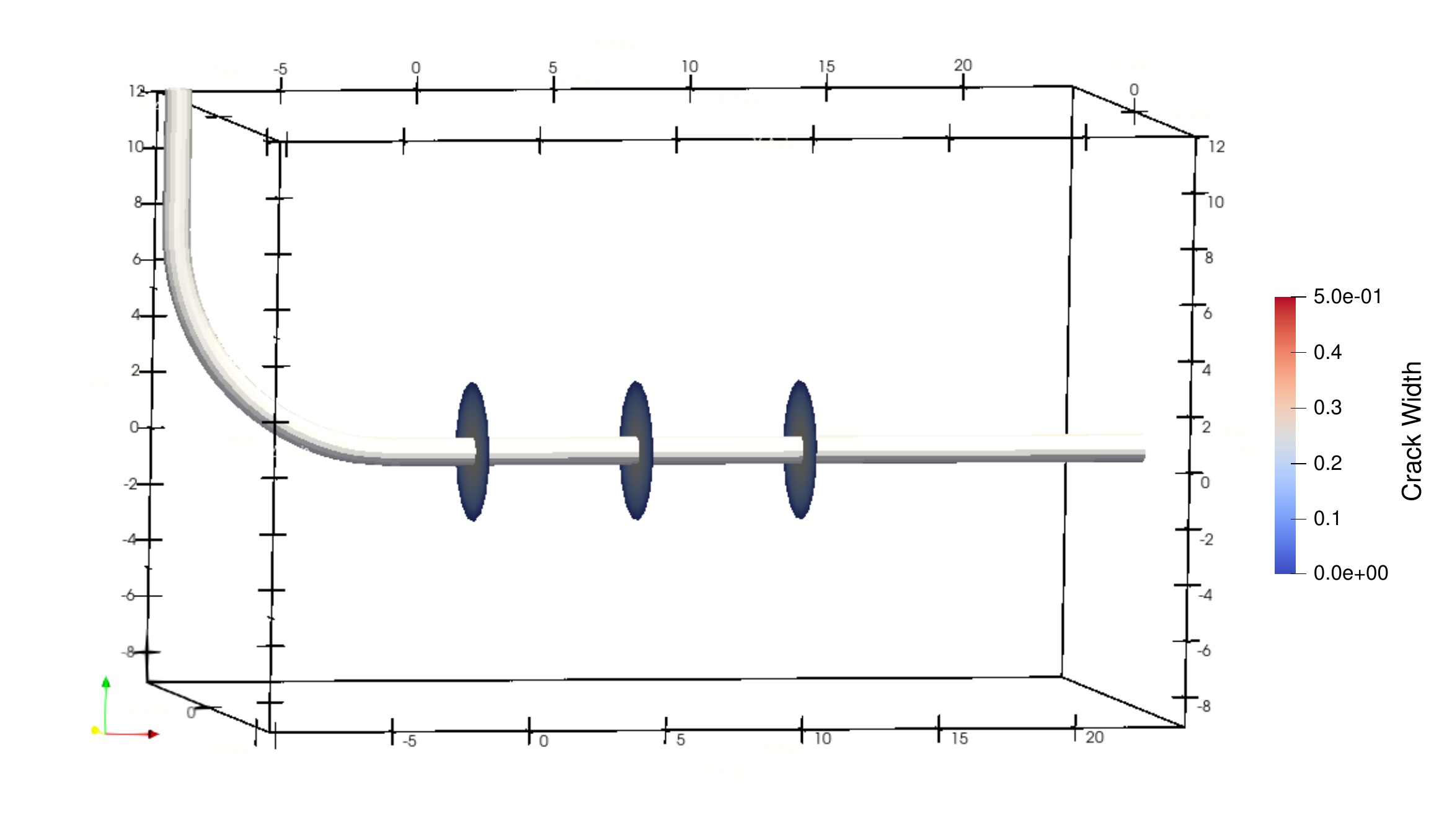}
		\caption{Case 2: Step 1}
		\label{fig:c2_s1}
	\end{subfigure}
	\hfill
	\begin{subfigure}[b]{0.32\textwidth}
		\centering
		\includegraphics[width=\textwidth]{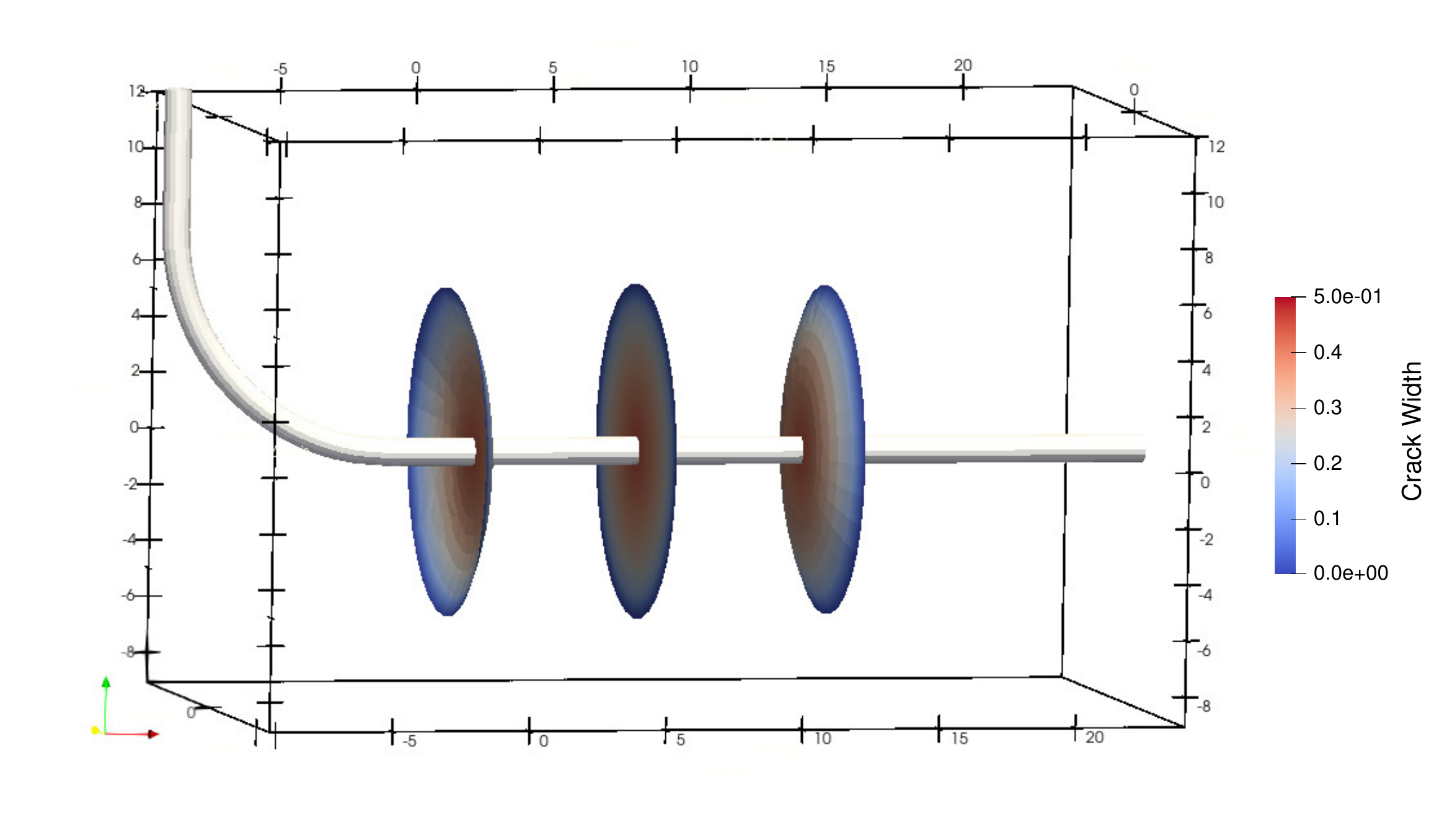}
		\caption{Case 2: Step 65}
		\label{fig:c2_s65}
	\end{subfigure}
	\hfill
	\begin{subfigure}[b]{0.32\textwidth}
		\centering
		\includegraphics[width=\textwidth]{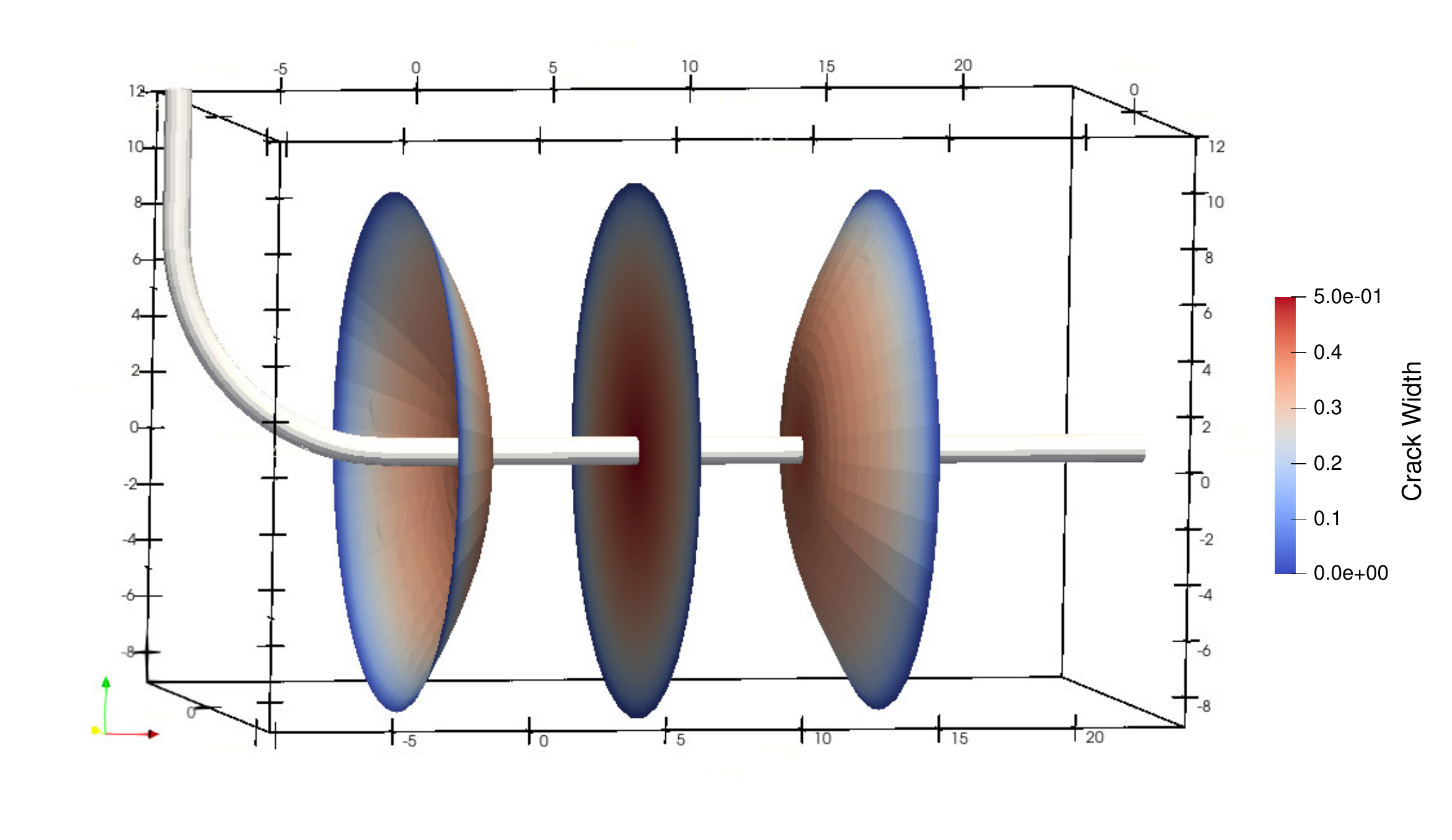}
		\caption{Case 2: Step 130}
		\label{fig:c2_s130}
	\end{subfigure}
	
	\caption{Evolution of 3D non-planar fracture trajectories. The top row demonstrates severe stress shadowing and outward curvature for a 3m interval (Case 1). The bottom row shows reduced interference for a 6m interval (Case 2).}
	\label{fig:spacing_propagation}
\end{figure*}

The temporal evolution of the fracture geometries is captured at steps 1, 65, and 130, as illustrated in Fig.~\ref{fig:spacing_propagation}. In Case 1 (Figs.~\ref{fig:c1_s1}--\ref{fig:c1_s130}), the tight 3-meter spacing generates a substantial compressive stress shadow that envelopes the central fracture. This induced compression restricts the central fracture's aperture and severely retards its longitudinal growth. Consequently, to minimize the elastic strain energy required to open against this elevated local stress, the two outer fractures are strongly repelled. They curve significantly outward, diverging away from the central fracture's plane. 

Conversely, in Case 2 (Figs.~\ref{fig:c2_s1}--\ref{fig:c2_s130}), doubling the spacing to 6 meters substantially attenuates the mechanical interference. The induced compressive stresses dissipate over the larger spatial domain. While the outer fractures still exhibit a subtle outward deflection due to residual shadowing, the overall propagation trajectories remain far more planar, and the central fracture is able to develop with a geometry highly comparable to its outer neighbors.

\subsubsection{Influence of Fluid Rheology}

To investigate the impact of fluid rheology on multi-stage interaction, Case 3 is introduced. This case maintains the highly constrained 3-meter spacing of Case 1 but replaces the low-viscosity Newtonian fluid with a highly viscous, shear-thinning power-law fluid. The power-law consistency index is set to $K = 0.2$~Pa$\cdot$s$^n$, and the flow behavior index is $n = 0.7$. 

\begin{figure*}[htbp]
	\centering
	\begin{subfigure}[b]{0.32\textwidth}
		\centering
		\includegraphics[width=\textwidth]{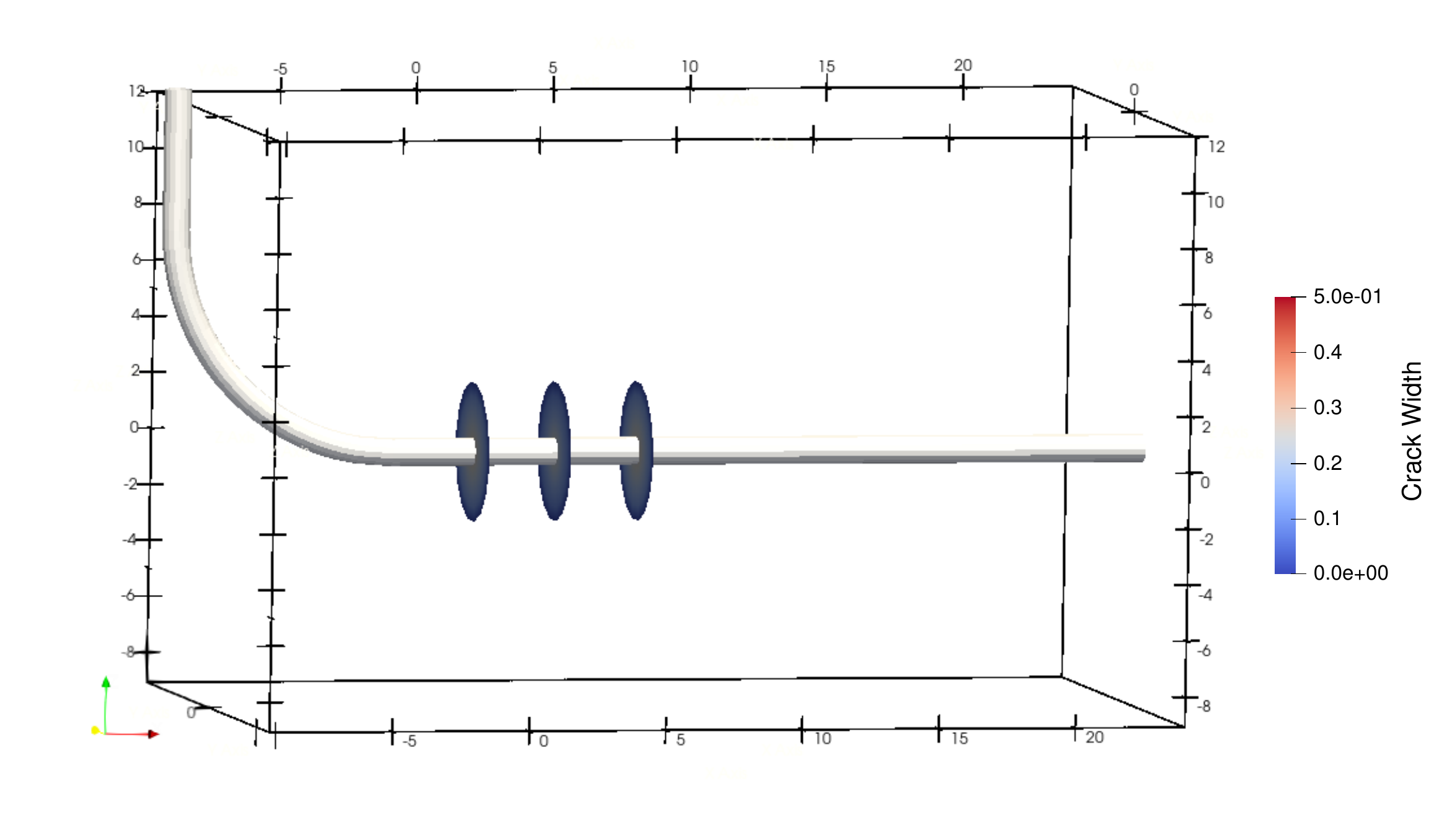}
		\caption{Case 3: Step 1}
		\label{fig:c3_s1}
	\end{subfigure}
	\hfill
	\begin{subfigure}[b]{0.32\textwidth}
		\centering
		\includegraphics[width=\textwidth]{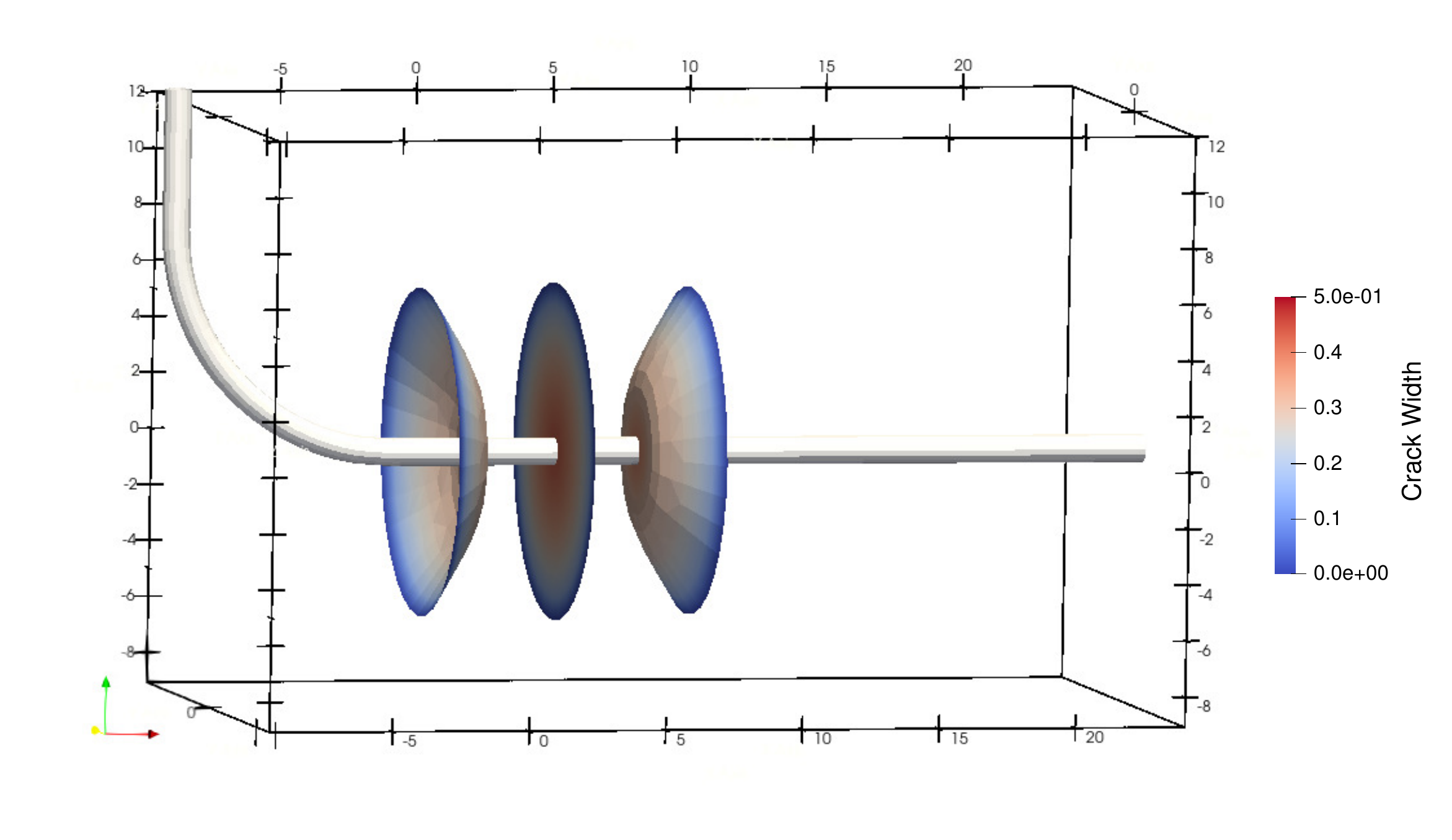}
		\caption{Case 3: Step 65}
		\label{fig:c3_s65}
	\end{subfigure}
	\hfill
	\begin{subfigure}[b]{0.32\textwidth}
		\centering
		\includegraphics[width=\textwidth]{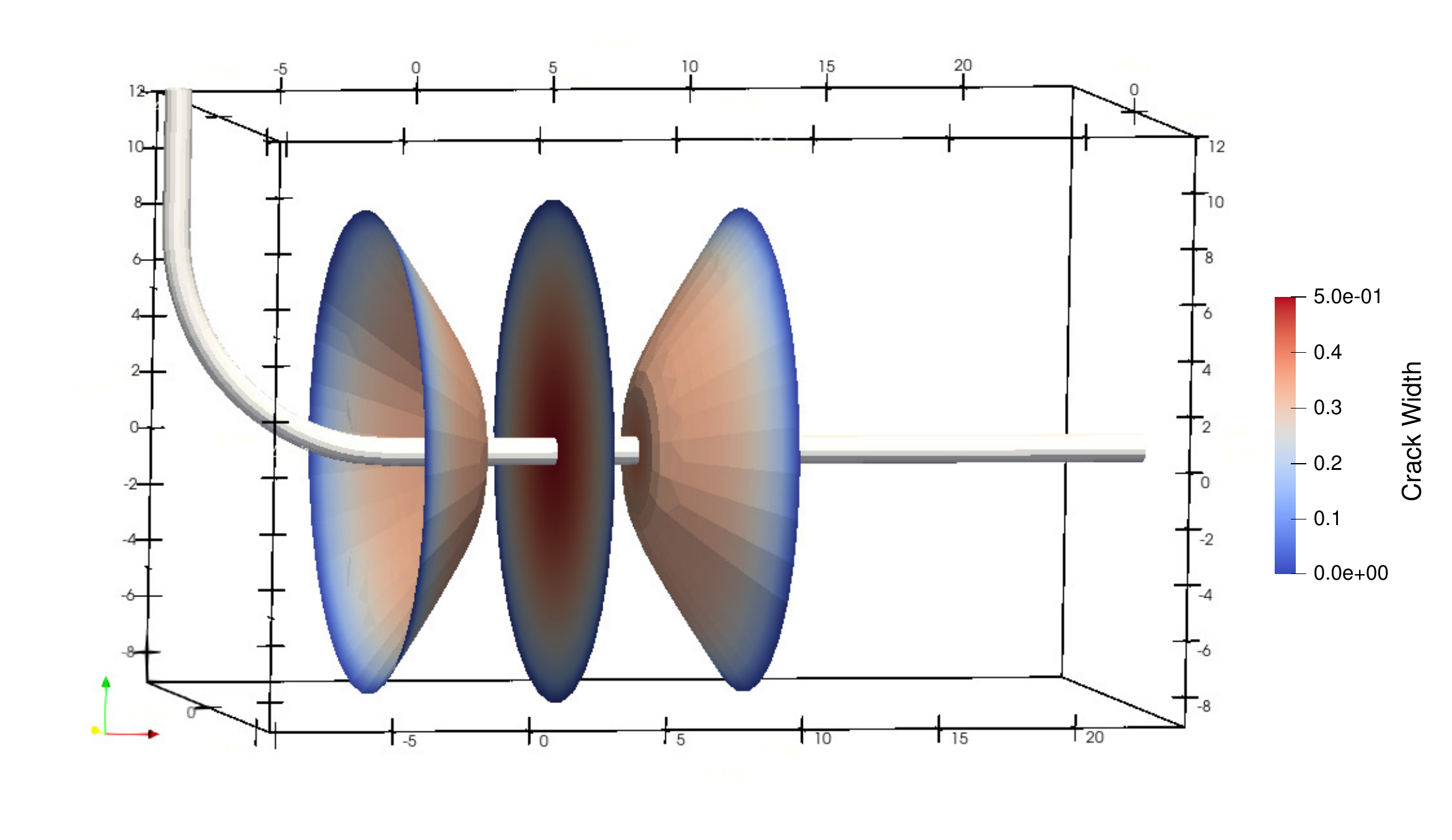}
		\caption{Case 3: Step 130}
		\label{fig:c3_s130}
	\end{subfigure}
	\caption{Propagation sequence for Case 3 utilizing a power-law fracturing fluid ($K=0.2$, $n=0.7$) at a 3m interval. While the internal pressure and fracture width are significantly larger due to fluid friction, the macroscopic trajectory remains nearly identical to the Newtonian equivalent.}
	\label{fig:rheology_propagation}
\end{figure*}

The propagation sequence for Case 3 is presented in Fig.~\ref{fig:rheology_propagation}. A comparative analysis between Case 1 (Newtonian) and Case 3 (Power-law) reveals two distinct physical responses. First, it is prominently observed that the fracture aperture (width) in Case 3 is significantly larger than in Case 1. Because the power-law fluid is inherently more viscous, it generates greater frictional resistance within the fracture channel. To maintain the mandated constant injection rate of $0.01$~m$^3$/s, the fluid pressure inside the fracture must increase substantially. This elevated internal net pressure directly translates to a wider fracture opening displacement.

Second, despite the marked increase in fracture width and internal pressure, the macroscopic turning trajectories of the fractures in Case 3 remain almost identical to those in Case 1. The fluid viscosity profile predominantly governs the internal pressure gradients and aperture dilations, rather than the macro-scale interaction paths. This observation confirms that under conditions of equal and constant injection rates, the spatial deflection of the fractures is overwhelmingly dominated by the far-field elastic stress shadow rather than fluid rheology. This physical consistency aligns with the earlier numerical observations reported by Castonguay et al. (2013)\ \cite{Castonguay2013}, which validates the HyFrac.fun platform's capability to resolve complex rheological and geomechanical effects without introducing spurious trajectory artifacts.

\subsubsection{Integrated Production Analysis}

A distinct advantage of the HyFrac.fun cloud architecture is the automated programmatic handoff of the final complex 3D fracture meshes directly into the steady-state Darcy production solver. By eliminating the need for manual mesh conversion, the platform allows for rapid lifecycle evaluations. For the production phase, the homogeneous reservoir is assigned a permeability of $k = 10$~mD, and the resident oil viscosity is set to $\mu_{res} = 250$~cP. A constant pressure drawdown of $\Delta P = 30$~MPa is applied at the wellbore perforations relative to the far-field reservoir boundary.

\begin{figure*}[htbp]
	\centering
	\begin{subfigure}[b]{0.32\textwidth}
		\centering
		\includegraphics[width=\textwidth]{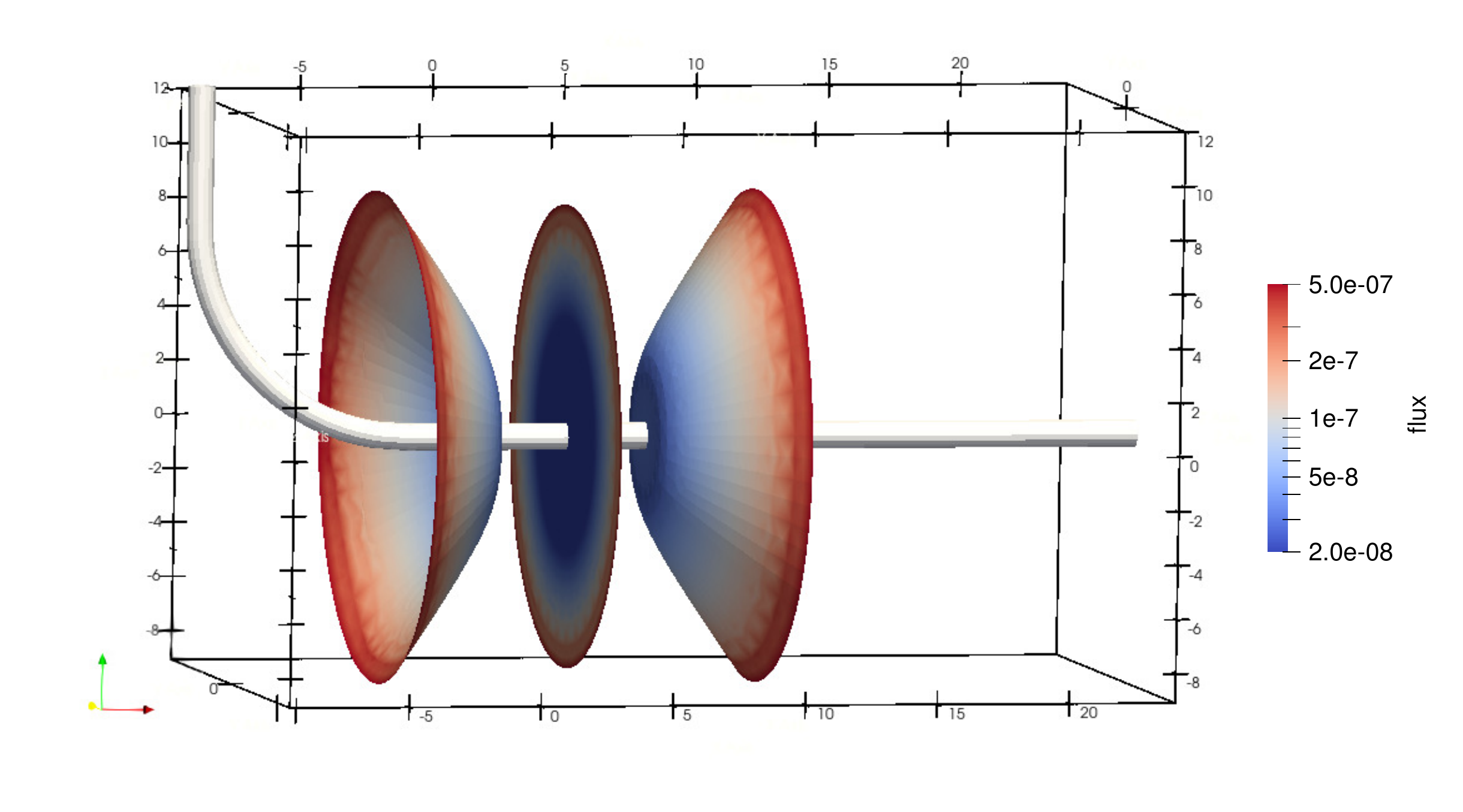}
		\caption{Case 1 Production}
		\label{fig:prod_c1}
	\end{subfigure}
	\hfill
	\begin{subfigure}[b]{0.32\textwidth}
		\centering
		\includegraphics[width=\textwidth]{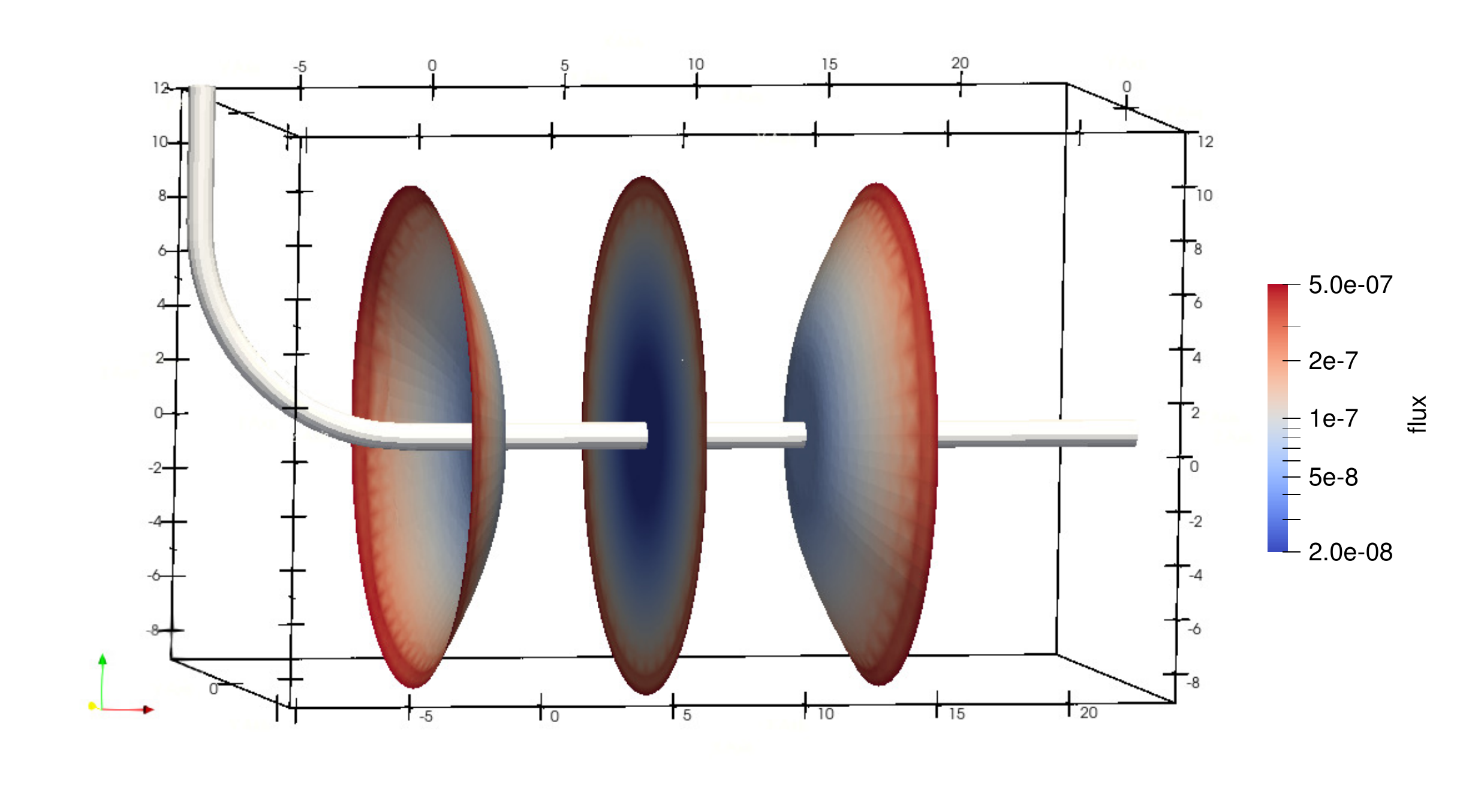}
		\caption{Case 2 Production}
		\label{fig:prod_c2}
	\end{subfigure}
	\hfill
	\begin{subfigure}[b]{0.32\textwidth}
		\centering
		\includegraphics[width=\textwidth]{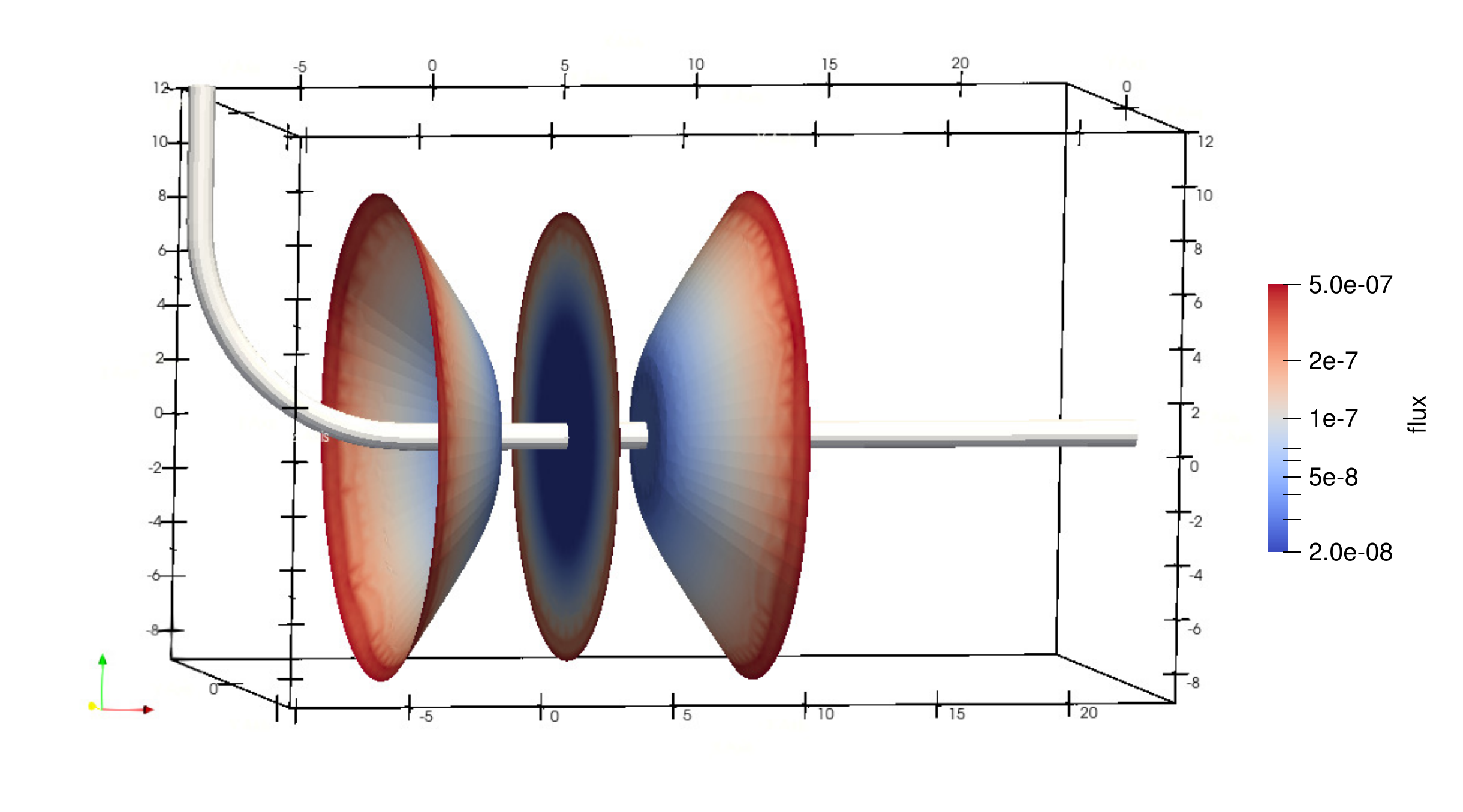}
		\caption{Case 3 Production}
		\label{fig:prod_c3}
	\end{subfigure}
	\caption{Steady-state flux distributions mapping the productivity of the final fracture networks. The intense fluidic shielding of the central fracture in the 3m configurations (a, c) is visually relieved in the wider 6m configuration (b).}
	\label{fig:production_comparison}
\end{figure*}

The steady-state production fluxes mapping the drainage efficiency for all three cases are visualized in Fig.~\ref{fig:production_comparison} (detailed in Figs.~\ref{fig:prod_c1}--\ref{fig:prod_c3}). The corresponding individual drawout rates are quantitatively summarized in Table~\ref{tab:production_rates}. 

The production analysis uncovers a double shadow phenomenon that connects the stimulation physics to the production physics through the fracture geometry. The same mechanical stress shadow that suppressed the central fracture's growth during propagation also shields it from reservoir drainage during production: the two outer fractures deplete the adjacent matrix, leaving the inner fracture surrounded by a pressure-depleted zone. The production analysis illustrates a fluidic pressure shadow effect that fundamentally mirrors the mechanical stress shadow observed during the stimulation phase. In Case 1 (3m interval), the inner fracture is heavily shielded from the surrounding reservoir by the drainage volumes of the two outer fractures. This flow interference yields an inner drawout rate $2.52 \times 10^{-5}$~m$^3$/s that is severely compromised, which is less than half that of the unshielded outer fractures. 

Conversely, expanding the stage spacing to 6m in Case 2 significantly mitigates this fluidic shielding. The increased separation allows the central fracture to access a larger, un-depleted matrix volume, increasing its individual contribution by roughly $46\%$ to $3.67 \times 10^{-5}$~m$^3$/s and boosting the total well productivity. The physically significant result is provided by Case 3. Despite the power-law fluid generating a substantially larger fracture aperture which is a consequence of higher viscous resistance in the channel flow, the macroscopic fracture trajectories and the resulting steady-state production rates are virtually identical to Case 1. This demonstrates that at equal injection rates, the stress-shadow-controlled fracture geometry is the primary determinant of long-term production efficiency, not the fracturing fluid rheology. The fracture width, which is rheology-dependent, governs conductance but not the spatial distribution of the drainage network; it is the latter that dominates production in tight formations.

These integrated results demonstrate the platform's utility in optimizing unconventional completion designs. By rapidly quantifying the coupled impact of mechanical and fluidic interference, reservoir engineers can iteratively evaluate designs to locate the economic optimum between multi-stage cluster density and individual fracture efficiency.

\begin{table*}[htbp]
	\caption{Steady-state drawout rates (m$^3$/s) for the three simulated fracture networks. The data quantitatively demonstrates the intense production shielding effect imposed on the central fracture in tightly spaced clusters.}
	\label{tab:production_rates}
	\centering
	\begin{tabular}{l c c c}
		\hline\hline
		Configuration & Fracture 1 (Outer) & Fracture 2 (Inner) & Fracture 3 (Outer) \\ [0.5ex] 
		\hline
		Case 1 (3m, Newtonian) & $5.38 \times 10^{-5}$ & $2.52 \times 10^{-5}$ & $5.38 \times 10^{-5}$ \\ 
		Case 2 (6m, Newtonian) & $5.40 \times 10^{-5}$ & $3.67 \times 10^{-5}$ & $5.42 \times 10^{-5}$ \\ 
		Case 3 (3m, Power-Law) & $5.38 \times 10^{-5}$ & $2.34 \times 10^{-5}$ & $5.38 \times 10^{-5}$ \\ 
		\hline\hline
	\end{tabular}
\end{table*}

\section{Conclusions}
\label{sec:conclusions}

This paper presents HyFrac.fun, a cloud-native platform that advances the 3D hydraulic fracture propagation framework of Rungamornrat et al. (2005)~\cite{Rungamornrat2005} and the production solver of Hu and Mear (2022)~\cite{Hu2022}.The specific improvements made over these prior works, which are fully documented adaptive remeshing, OpenMP-parallelized 3D stiffness assembly, dynamic solver-switching, and the first application of the production engine to evolved non-planar propagation geometries, are unified through a cloud-native lifecycle pipeline whose contributions are summarized below.

\noindent 1. Operator-based lifecycle integration: A structural analysis of the two governing operator systems reveals that the hydraulic fracture propagation and steady-state production problems are governed by structurally isomorphic abstract operator families $\{\mathcal{A}, \mathcal{B}, \mathcal{C}, \mathcal{R}\}$. This isomorphism enables the automated zero-conversion transfer of the evolved 3D fracture mesh and physical field state from the propagation solver to the production solver, realizing HyFrac.fun as an integrated lifecycle simulation platform that connects non-planar 3D multi-stage hydraulic fracture propagation directly to steady-state Darcy production analysis on the resulting evolved geometry.

\noindent 2. Detailed adaptive remeshing algorithm: The adaptive remeshing procedure employed by the propagation engine, which comprises a four-case element splitting and merging decision tree, a characteristic-length-driven mesh quality control scheme, and a localized solution-field projection with singularity regularization for newly created crack-tip nodes, is documented in full detail in this paper for the convenient implementation. Although this algorithm underpins the results~\cite{Rungamornrat2005,hu2024efficient}, its implementation has not been previously published in the open literature.

\noindent 3. Performance engineering for non-planar multi-stage fractures: The incremental stiffness update strategy, cache-optimized element-renumbering permutation, and hybrid OpenMP parallelization of Mood (2019)\cite{mood2019coupled} and Hu et al. (2024)~\cite{hu2024efficient} are extended to the fully non-planar arbitrarily curved fracture networks considered here. A dynamic solver-switching algorithm is introduced to ensure fault-tolerant operation under the extreme Jacobian ill-conditioning arising from violent stress-shadow interactions between adjacent fractures.

\noindent 4. Cloud-native service-oriented architecture: A three-tier architecture, i.e. FastAPI asynchronous backend, TRAME/VTK server-side rendering service, and lightweight HTML5 front-end, orchestrates the entire lifecycle as a single reproducible server-side workflow, delivering interactive 3D visualization through a standard browser without requiring local HPC resources or proprietary visualization software.

Through rigorous benchmark validations and complex multi-stage interaction scenarios, the platform demonstrated its capacity to accurately capture critical geomechanical phenomena, such as stress-induced trajectory deviations and rheology-driven aperture dilations. The fully automated propagation-to-production workflow provided quantitative insights into how mechanical interference during the stimulation phase dictates long-term reservoir drainage, revealing a pronounced ``pressure shadow'' that impairs the drawout efficiency of clustered fractures. Ultimately, HyFrac.fun democratizes access to advanced computational geomechanics, providing a scalable blueprint for the future of simulation-as-a-service.

\begin{acknowledgments}

The authors wish to acknowledge the support through Shanghai Pujiang Program (22PJD074) and Open Fund of Key Laboratory of Petroleum Resources Exploration and Evaluation, Gansu Province (KLPREEGS-2024-10).
\end{acknowledgments}

\section{Data Availability Statement}

The source code for the cloud orchestration platform, including the web frontend, FastAPI middleware, and remote visualization service, is available at \url{https://github.com/JINGHU-UT/HF_cloud} under the MIT License.

\appendix

\bibliography{aipsamp.bib}

@PREAMBLE{
	"\providecommand{\noopsort}[1]{}" 
	# "\providecommand{\singleletter}[1]{#1}%" 
}

@ARTICLE{Montgomery2010,
	author       = "C. T. Montgomery and M. B. Smith",
	title        = "Hydraulic Fracturing: History of an Enduring Technology",
	journal      = "Journal of Petroleum Technology",
	year         = "2010",
	volume       = "62",
	number       = "12",
	pages        = "26--40",
}

@techreport{Gallegos2015,
	author       = "T. J. Gallegos and B. A. Varela",
	title        = "Trends in hydraulic fracturing distributions and treatment fluids, additives, proppants, and water volumes applied to wells drilled in the United States from 1947 through 2010: Data analysis and comparison to the literature",
	year         = "2015",
	institution = "US Geological Survey"
}

@ARTICLE{Clark1949,
	author       = "J. B. Clark",
	title        = "A Hydraulic Process for Increasing the Productivity of Wells",
	journal      = "Journal of Petroleum Technology",
	year         = "1949",
	volume       = "1",
	pages        = "1--8",
}

@ARTICLE{Hubbert1957,
	author       = "M. K. Hubbert and D. G. Willis",
	title        = "Mechanics of Hydraulic Fracturing",
	journal      = "Transactions of the AIME",
	year         = "1957",
	volume       = "210",
	pages        = "153--168",
}

@BOOK{Valko1995,
	author       = "P. Valko and M. J. Economides",
	title        = "Hydraulic Fracture Mechanics",
	year         = "1995",
	publisher    = "John Wiley \& Sons",
	address      = "Wiley Chichester",
}

@BOOK{Economides2000,
	author       = "M. J. Economides and K. G. Nolte",
	title        = "Reservoir Stimulation",
	year         = "2000",
	publisher    = "Wiley",
	address      = "New York",
}

@BOOK{Yew1997,
	author       = "C. H. Yew and X. Weng",
	title        = "Mechanics of Hydraulic Fracturing",
	year         = "2014",
	publisher    = "Gulf Professional Publishing",
	address      = "Houston",
}

@ARTICLE{Warpinski1987,
	author       = "N. R. Warpinski and L. W. Teufel",
	title        = "Influence of Geologic Discontinuities on Hydraulic Fracture Propagation",
	journal      = "Journal of Petroleum Technology",
	year         = "1987",
	volume       = "39",
	number       = "02",
	pages        = "209--220",
}

@ARTICLE{Daneshy1973,
	author       = "A. A. Daneshy",
	title        = "On the Design of Vertical Hydraulic Fractures",
	journal      = "Journal of Petroleum Technology",
	year         = "1973",
	volume       = "25",
	number       = "01",
	pages        = "83--97",
}

@ARTICLE{Cipolla2008,
	author       = "C. L. Cipolla and E. P. Lolon and J. C. Erdle and B. Rubin",
	title        = "Reservoir Modeling in Shale-Gas Reservoirs",
	journal      = "SPE Reservoir Evaluation \& Engineering",
	year         = "2010",
	volume       = "13",
	number       = "04",
	pages        = "638--653",
}

@ARTICLE{Mayerhofer2010,
	author       = "M. J. Mayerhofer and E. P. Lolon and N. R. Warpinski and C. L. Cipolla and D. Walser and C. M. Rightmire",
	title        = "What is Stimulated Reservoir Volume?",
	journal      = "SPE Production \& Operations",
	year         = "2010",
	volume       = "25",
	number       = "01",
	pages        = "89--98",
}

@inproceedings{Olson2008,
	author       = "J. E. Olson",
	title        = "Multi-fracture propagation modeling: Applications to hydraulic fracturing in shales and tight gas sands",
	booktitle      = "ARMA US Rock Mechanics/Geomechanics Symposium",
	year         = "2008",
	pages        = "ARMA--08",
}

@ARTICLE{Castonguay2013,
	author       = "S. T. Castonguay and M. E. Mear and R. H. Dean and J. H. Schmidt",
	title        = "Predictions of the growth of multiple interacting hydraulic fractures in three dimensions",
	journal      = "SPE Annual Technical Conference and Exhibition",
	year         = "2013",
	pages={D031S048R001},
	organization={SPE}
}

@ARTICLE{Adachi2007,
	author       = "J. Adachi and E. Siebrits and A. Peirce and J. Desroches",
	title        = "Computer simulation of hydraulic fractures",
	journal      = "International Journal of Rock Mechanics and Mining Sciences",
	year         = "2007",
	volume       = "44",
	number       = "5",
	pages        = "739--757",
}

@ARTICLE{Weng2011,
	author       = "X. Weng and O. Kresse and C. Cohen and R. Wu and H. Gu",
	title        = "Modeling of hydraulic-fracture-network propagation in a naturally fractured formation",
	journal      = "SPE Production \& Operations",
	year         = "2011",
	volume       = "26",
	number       = "04",
	pages        = "368--380",
}

@ARTICLE{Geertsma1969,
	author       = "J. Geertsma and R.Haafkens",
	title        = "A comparison of the theories for predicting width and extent of vertical hydraulically induced fractures",
	journal      = "Journal of energy resources technology",
	year         = "1979",
	volume       = "101",
	number       = "1",
	pages        = "8--19",
}

@inproceedings{Khristianovic1955,
	author       = "S. A. Khristianovic and Y. P. Zheltov",
	title        = "Formation of vertical fractures by means of highly viscous fluids",
	journal      = "Proceedings of the 4th World Petroleum Congress",
	year         = "1955",
	pages        = "579-586",
}

@ARTICLE{Perkins1961,
	author       = "T. K. Perkins and L. R. Kern",
	title        = "Widths of Hydraulic Fractures",
	journal      = "Journal of petroleum technology",
	year         = "1961",
	volume       = "13",
	number       = "09",
	pages        = "937--949",
}

@ARTICLE{Nordgren1972,
	author       = "R. P. Nordgren",
	title        = "Propagation of a Vertical Hydraulic Fracture",
	journal      = "Society of petroleum engineers journal",
	year         = "1972",
	volume       = "12",
	number       = "04",
	pages        = "306--314",
}

@ARTICLE{settari1986development,
	author       = "A. Settari and M. P. Cleary",
	title        = "Development and testing of a pseudo-three-dimensional model of hydraulic fracture geometry",
	journal      = "SPE Production Engineering",
	year         = "1986",
	volume       = "1",
	number       = "06",
	pages        = "449--466",
	publisher    = "OnePetro"
}

@INPROCEEDINGS{Cleary1980,
	author       = "M. P. Cleary",
	title        = "Comprehensive design formulae for hydraulic fracturing",
	booktitle    = "Proc. SPE Annual Technical Conference and Exhibition",
	year         = "1980",
	pages        = "SPE--9259",
}

@INPROCEEDINGS{Meyer1986,
	author       = "B. R. Meyer",
	title        = "Design Formulae for {2-D} and {3-D} Vertical Hydraulic Fractures: Model Comparison and Parametric Studies",
	booktitle      = "SPE Unconventional Resources Conference/Gas Technology Symposium",
	year         = "1986",
	pages        = "SPE--15240",
}

@ARTICLE{Lecampion2018,
	author       = "B. Lecampion and A. Bunger and X. Zhang",
	title        = "Numerical methods for hydraulic fracture propagation: A review of recent trends",
	journal      = "Journal of natural gas science and engineering",
	year         = "2018",
	volume       = "49",
	pages        = "66--83",
}

@ARTICLE{Gordeliy2013,
	author       = "E. Gordeliy and A. Peirce",
	title        = "Implicit level set schemes for modeling hydraulic fractures using the XFEM",
	journal      = "Computer Methods in Applied Mechanics and Engineering",
	year         = "2013",
	volume       = "266",
	pages        = "125--143",
}

@ARTICLE{miehe2010phase,
	author       = "C. Miehe and M. Hofacker and F. Welschinger",
	title        = "A phase field model for rate-independent crack propagation: Robust algorithmic implementation based on operator splits",
	journal      = "Computer Methods in Applied Mechanics and Engineering",
	year         = "2010",
	volume       = "199",
	number       = "45-48",
	pages        = "2765--2778",
	publisher    = "Elsevier"
}

@ARTICLE{Bourdin2008,
	author       = "B. Bourdin and G. A. Francfort and J. J. Marigo",
	title        = "The variational approach to fracture",
	journal      = "Journal of elasticity",
	year         = "2008",
	volume       = "91",
	number       = "1",
	pages        = "5--148",
}

@ARTICLE{Bonnet1998,
	author       = "M. Bonnet and G. Maier and G. Polizzotto",
	title        = "Symmetric Galerkin boundary element methods",
	journal      = "Applied Mechanics Reviews",
	year         = "1998",
	pages        = "669--704",
}

@BOOK{sutradhar2008symmetric,
	author       = "A. Sutradhar and G. Paulino and L. J. Gray",
	title        = "Symmetric Galerkin boundary element method",
	year         = "2008",
	publisher    = "Springer Science \& Business Media"
}

@ARTICLE{Phan2003,
	author       = "A. V. Phan and J. Napier and L. Gray and T. Kaplan",
	title        = "Symmetric-Galerkin BEM simulation of fracture with frictional contact",
	journal      = "International journal for numerical methods in engineering",
	year         = "2003",
	volume       = "57",
	number       = "6",
	pages        = "835--851",
}

@INPROCEEDINGS{Rungamornrat2005,
	author       = "J. Rungamornrat and M. F. Wheeler and M. E. Mear",
	title        = "A numerical technique for simulating nonplanar evolution of hydraulic fractures",
	booktitle = "SPE Annual Technical Conference and Exhibition",
	year         = "2005",
	pages        = "SPE--96968",
}

@ARTICLE{Ganis2014,
	author       = "B. Ganis and M. Mear and A. Sakhaee-Pour and M. F. Wheeler and T. Wick",
	title        = "Modeling fluid injection in fractures with a reservoir simulator coupled to a boundary element method",
	journal      = "Computational Geosciences",
	year         = "2014",
	volume       = "18",
	number       = "5",
	pages        = "613--624",
}

@ARTICLE{Hu2025,
	author       = "J. Hu and M. E. Mear",
	title        = "A computational framework for well production simulation: Coupling transient Darcy flow and channel flow by SGBEM--FEM",
	journal      = "Computer Methods in Applied Mechanics and Engineering",
	year         = "2025",
	volume       = "433",
	pages        = "117491",
}

@ARTICLE{hu2024efficient,
	author       = "J. Hu and C. G. Mood and and M. E. Mear",
	title        = "An efficient computational framework for height-contained growing and intersecting hydraulic fracturing simulation via SGBEM--FEM",
	journal      = "Computer Methods in Applied Mechanics and Engineering",
	year         = "2024",
	volume       = "419",
	pages        = "116653",
}

@INPROCEEDINGS{omidi2015adaptive,
	author       = "O. Omidi and R. Abedi and S. Enayatpour",
	title        = "An adaptive meshing approach to capture hydraulic fracturing",
	booktitle = "ARMA US Rock Mechanics/Geomechanics Symposium",
	year         = "2015",
	pages        = "ARMA--2015",
}

@ARTICLE{cotterell1980slightly,
	author       = "B. Cotterell and J. R. Rice",
	title        = "Slightly curved or kinked cracks",
	journal      = "International journal of fracture",
	year         = "1980",
	volume       = "16",
	number       = "2",
	pages        = "155--169",
	publisher    = "Springer"
}

@BOOK{mood2019coupled,
	author       = "C. G. Mood",
	title        = "Coupled SGBEM-FEM for efficient simulation of height-contained hydraulic fractures",
	year         = "2019",
	publisher    = "The University of Texas at Austin"
}

@BOOK{chandra2001parallel,
	author       = "R. Chandra",
	title        = "Parallel programming in OpenMP",
	year         = "2001",
	publisher    = "Morgan kaufmann"
}

@ARTICLE{Dagum1998,
	author       = "L. Dagum and R. Menon",
	title        = "OpenMP: an industry standard API for shared-memory programming",
	journal      = "IEEE computational science and engineering",
	year         = "1998",
	volume       = "5",
	number       = "1",
	pages        = "46--55",
	publisher    = "IEEE"
}

@ARTICLE{heidbach2020manual,
	author       = "O. Heidbach and M. Ziegler and D. Stromeyer",
	title        = "Manual of the tecplot 360 add-on geostress v2. 0",
	year         = "2020",
	publisher    = "GFZ German Research Centre for Geosciences"
}

@ARTICLE{Ahrens2005,
	author       = "J. Ahrens and B. Geveci and C. Law",
	title        = "Paraview: An end-user tool for large data visualization",
	journal      = "The visualization handbook",
	year         = "2005",
	volume       = "717",
	number       = "8"
}

@BOOK{Schroeder2006,
	author       = "W. Schroeder and K. Martin and B. Lorensen",
	title        = "The visualization toolkit an object-oriented approach to 3D graphics 4th edition",
	year         = "2006",
	publisher    = "Kitware"
}

@ARTICLE{Armbrust2010,
	author       = "M. Armbrust and A. Fox and R. Griffith and A. D. Joseph and R. Katz and A. Konwinski and G. Lee and D. Patterson and A. Rabkin and I. Stoica and M. Zaharia",
	title        = "A view of cloud computing",
	journal      = "Communications of the ACM",
	year         = "2010",
	volume       = "53",
	number       = "4",
	pages        = "50--58",
}

@ARTICLE{Buyya2009,
	author       = "R. Buyya and C. S. Yeo and S. Venugopal and J. Broberg and I. Brandic",
	title        = "Cloud computing and emerging IT platforms: Vision, hype, and reality for delivering computing as the 5th utility",
	journal      = "Future Generation computer systems",
	year         = "2009",
	volume       = "25",
	number       = "6",
	pages        = "599--616",
}

@INPROCEEDINGS{Foster2008,
	author       = "I. Foster and Y. Zhao and I. Raicu and S. Lu",
	title        = "Cloud Computing and Grid Computing {360}-Degree Look",
	booktitle    = "2008 grid computing environments workshop",
	year         = "2008",
	pages        = "1--10",
}

@INPROCEEDINGS{Zhao2007,
	author       = "Y. Zhao and C. Hu and Y. Huang and D. Ma",
	title        = "Collaborative visualization of large scale datasets using web services",
	booktitle      = "Second International Conference on Internet and Web Applications and Services (ICIW'07)",
	year         = "2007",
	pages        = "62",
}

@ARTICLE{Benlian2010,
	author       = "A. Benlian and T. Hess",
	title        = "Opportunities and risks of software-as-a-service: Findings from a survey of IT executives",
	journal      = "Decision support systems",
	year         = "2011",
	volume       = "52",
	number       = "1",
	pages        = "232--246",
}

@BOOK{lubanovic2023fastapi,
	author       = "B. Lubanovic",
	title        = "FastAPI",
	year         = "2023",
	publisher    = "O'Reilly Media, Inc."
}

@MISC{Trame2021,
	author       = "{Kitware}",
	title        = "Trame",
	howpublished = "\url{https://kitware.github.io/trame/}",
	year         = "2021",
}

@ARTICLE{Sneddon1946,
	author       = "I. N. Sneddon",
	title        = "The distribution of stress in the neighbourhood of a crack in an elastic solid",
	journal      = "Proc. R. Soc. Lond. A",
	year         = "1946",
	volume       = "187",
	pages        = "229--260"
}

@ARTICLE{Hu2022,
	author       = "J. Hu and M. E. Mear",
	title        = "A computational framework for well production simulation: Coupling steady state {Darcy} flow and channel flow by {SGBEM-FEM}",
	journal      = "Comput. Methods Appl. Mech. Engrg.",
	year         = "2022",
	volume       = "399",
	pages        = "115300"
}

@INPROCEEDINGS{palmer1983numerical,
	author       = "I. D. Palmer and H. B. Carroll Jr",
	title        = "Numerical solution for height and elongated hydraulic fractures",
	booktitle    = "SPE Rocky Mountain Petroleum Technology Conference/Low-Permeability Reservoirs Symposium",
	year         = "1983",
	pages        = "SPE--11627",
	organization = "SPE"
}

@ARTICLE{carter1957derivation,
	author       = "R. D. Carter and G. C. Howard and C. R. Fast and others",
	title        = "Derivation of the general equation for estimating the extent of the fractured area",
	journal      = "Appendix I of drilling and production practice/Ed. by GC Howard, CR Fast. NY: American Petroleum Institute",
	year         = "1957",
	pages        = "261--270"
}
\end{document}
%